\documentclass[%
 aip,
 amsmath,amssymb,
 reprint,%
]{revtex4-2}

\usepackage{graphicx}
\usepackage{dcolumn}
\usepackage{bm}
\usepackage{xcolor}
\usepackage{mathrsfs}

\usepackage[utf8]{inputenc}
\usepackage[T1]{fontenc}
\usepackage{mathptmx}
\usepackage{etoolbox}
\usepackage{ulem}

\def\v#1{{\boldsymbol{#1}}}  

\makeatletter
\def\@email#1#2{%
 \endgroup
 \patchcmd{\titleblock@produce}
  {\frontmatter@RRAPformat}
  {\frontmatter@RRAPformat{\produce@RRAP{*#1\href{mailto:#2}{#2}}}\frontmatter@RRAPformat}
  {}{}
}%
\makeatother
\begin{document}

\preprint{AIP/123-QED}

\title[Article]{Numerical-experimental estimation of the deformability of human red blood cells from rheometrical data}

\author{Naoki Takeishi}
 \altaffiliation[Authors to whom correspondence should be addressed: ]{takeishi.naoki.008@m.kyushu-u.ac.jp}
 \affiliation{Department of Mechanical Engineering, Kyushu University, 744 Motooka, Nishi-ku, Fukuoka 819-0395, Japan}

\author{Tomohiro Nishiyama}
 \affiliation{Department of Pure and Applied Physics, Kansai University, 3-3-35 Yamate-cho, Suita, Osaka 564-8680, Japan}

\author{Kodai Nagaishi}
 \affiliation{Department of Pure and Applied Physics, Kansai University, 3-3-35 Yamate-cho, Suita, Osaka 564-8680, Japan}

\author{Takeshi Nashima}
 \affiliation{OhnaTech Inc., 2-1-6 Sengen, Tsukuba, Ibraki 305-0047, Japan}

\author{Masako Sugihara-Seki}
\affiliation{Department of Pure and Applied Physics, Kansai University, 3-3-35 Yamate-cho, Suita, Osaka 564-8680, Japan}

\date{\today}

\begin{abstract}
The deformability of human red blood cells (RBCs),
which comprise almost 99\% of the cells in whole blood,
is largely related not only to pathophysiological blood flow but also to the levels of intracellular compounds.
Therefore, statistical estimates of the deformability of individual RBCs are of paramount importance in the clinical diagnosis of blood diseases.
Although the micro-scale hydrodynamic interactions of individual RBCs lead to non-Newtonian blood rheology,
there is no established method to estimate individual RBC deformability from the rheological data of RBC suspensions,
and the possibility of this estimation has not been proven.
To address this issue,
we conducted an integrated analysis of a model of the rheology of RBC suspensions,
coupled with macro-rheological data of human RBCs suspended in plasma. 
Assuming a non-linear curve of the relative viscosity of the suspensions as a function of the cell volume fraction,
the statistical average of the membrane shear elasticity was estimated for individual intact RBCs or hardened RBCs.
Both estimated values reproduced well the experimentally observed shear-thinning non-Newtonian behavior in these suspensions.
We hereby conclude that our complementary approach makes it possible to estimate the statistical average of individual RBC deformability from macro-rheological data obtained with usual rheometric tests.
\end{abstract}

\maketitle


\section{\label{sec:sec1}INTRODUCTION}

The human red blood cell (RBC) count is measured in millions per cubic millimeter (millions~mm$^{-3}$),
with normal values varying slightly from 3.9 to 5.1 millions~mm$^{-3}$ in women and 4.5 to 5.6 million~mm$^{-3}$ in men~\cite{Wakeman2007}.
Therefore, the flow behavior of individual RBCs,
which is a result of the interaction between a fluid and the RBC membrane,
is of significant importance in blood rheology.
To precisely diagnose blood diseases,
the mechanical properties of RBCs,
including shape recovery~\cite{Ito2017}, membrane elasticity~\cite{Saadat2020}, and clinical quality~\cite{Isiksacan2023},
have been intensively investigated in on-chip cell-manipulation systems designed based on the principle of fast, fine cell manipulation inside a microchannel~\cite{Ito2020}.
Any decrease in RBC deformability will affect flow resistance, tissue perfusion, and oxygenation~\cite{Parthasarathi1999}.
For instance, an approximately 16\% decrease in RBC deformability caused a 75\% increase in whole-flow resistance in isolated perfused rat hind limbs~\cite{Baskurt2004}.
The loss of RBC deformability has been observed in various hematologic disorders,
such as sickle cell disease~\cite{Chien1970JClinInvest, Manwani2013},
malaria~\cite{Safeukui2008, Suresh2005},
and type 2 diabetes mellitus~\cite{Agrawal2016},
and is correlated with impaired perfusion, increased blood viscosity, and an elevated likelihood of vaso-occlusion.
In addition, when RBCs lose their deformability,
they become more fragile and prone to hemolysis, which is the root cause of chronic hemolytic anemia in sickle cell disease~\cite{Connes2014}.
Furthermore, all living cells, including RBCs, are accurately functionalized in association with the sophisticated structures and metabolic energies generated by consuming adenosine triphosphate (ATP)~\cite{Eguchi1997, Forsyth2011}.
Thus, understanding cell deformability may allow for the estimation not only of the state of intracellular compounds,
but also of the levels of metabolic waste in organs, e.g., in~\citet{Teruya2021}.
Thus, the fast, precise estimation of cell deformability and viability is required for more accurate clinical diagnoses.

RBC deformability has been well investigated for over several decades~\cite{Chien1987}.
In more recently,
researchers successfully quantified single-cell deformability at high-throughput by exposing cells to a shear flow in a microfluidic channel,
e.g., in~\citet{Fregin2019, Toepfner2018, Otto2015},
where numerical-experimental integrated analysis was applied to estimate hydrodynamic stress acting on the deformed cell surface.
For instance, \citet{Toepfner2018} analyzed the size and stiffness of human blood cells without any labeling or separation at rates of $1000$ cells~sec$^{-1}$.
Along with these studies,
recent data-driven approaches may allow for the estimation of the mechanical properties of individual RBCs from experimentally measurable data, e.g., morphological features from single-cell microscope images~\cite{Lamoureux2022, Rizzuto2021}.
Since such a posteriori approaches are well suited to big data and artificial intelligence (AI),
they have recently attracted particular attention not only in hematological diagnosis~\cite{Guncar2018} but also in transfusion medicine~\cite{Isiksacan2023}.
Despite these efforts,
there is still no established methodology for estimating the statistical average of mechanical properties of individual RBCs from (macro-)rheological data,
which can be obtained with usual rheometric tests,
and the possibility of such estimations has not been proved.
Therefore our primary question in this study is {\it whether we can estimate the deformability of individual RBCs from the rheological data of RBC suspensions}.
In particular, we examine whether the membrane shear elasticity $G_s$,
as determined from the shear viscosity by means of experimental measurements,
is indeed comparable to the value obtained with a single-cell mechanical stretch test, e.g., in~\citet{Suresh2005}.
Since the 
fluid-membrane interactions are also affected by the contrast in viscosity between the cytoplasm ($\eta_1$) and plasma ($\eta_0$),
the interplay between a fluid and membrane is of paramount importance in estimating cell behavior.
The next question, therefore, is {\it how the viscosity ratio $\eta_1/\eta_0$ between the cytoplasm and plasma alters the relative membrane shear elasticity at the single-cell level}.

It is well known that human whole blood (WB) exhibits non-Newtonian shear-thinning behavior~\cite{Wells1961, Dintenfass1968} that is determined by RBC aggregation at low shear rates ($O(\dot\gamma) \leq 10^1$ s$^{-1}$) and by cell deformability at relatively large shear rates ($O(\dot\gamma) \geq 10^1$ s$^{-1}$)~\cite{Chien1970, Chien1975}.
A drastic initial decrease in shear viscosity for up to tens of seconds$^{-1}$ is accompanied by a breakdown of microstructures, the so-called rouleaux.
It is known that such clustered RBCs is caused by attractive forces between RBCs induced by macro-molecules in the plasma,
such as fibrinogen~\cite{Merrill1966}.
Further decreased WB viscosity 
for $O(\dot\gamma) \geq 10^1$ s$^{-1}$
relies only on the deformability 
of RBCs,
such as their membrane rotation and elongation towards the flow direction with the cells in an ellipsoidal shape,
the so-called tank-treading motion~\cite{Schmid-Schonbein1969, Fischer1978}.
Then, the blood viscosity reaches a constant value for $\geq 1000$ s$^{-1}$~~\cite{Chien1970, Chien1975}.
A previous experimental study showed that the critical shear rates (ranging $13-120$ s$^{-1}$, $38$ s$^{-1}$ at the volume fraction of $45$\%) required for complete cell disaggregation decreased as the RBC volume fractions (ranging $15$\%--$75$\%) increases~\cite{Snabre1987}.
Other studies have also shown that RBCs can be fluid-like objects that undergo a tank-treading motion when subjected to shear flow in the range of $\dot\gamma = 28-575$ s$^{-1}$~\cite{Fischer1978, Fischer1980}.
An experimental measurement, e.g., by~\citet{Chien1970} revealed that RBC deformation alone is sufficient to give rise to shear-thinning behavior.

Along with these experimental studies, 
numerical model analyses have successfully connected the aforementioned 
fluid-membrane interaction with suspension rheology.
A numerical simulation revealed the RBC (membrane) deformability alone as a factor of shear-thinning behavior even at the single-cell level~\cite{Omori2014}.
Limited numerical analyses successfully reproduced the shear-thinning behavior of a suspension of RBCs based on hydrodynamic interactions between a fluid and membrane~\cite{Fedosov2011, Lanotte2016, Takeishi2019}.
A more recent numerical study extended this knowledge to the viscoelasticity of RBC suspensions under oscillatory shear flow~\cite{Takeishi2024}.
The classification and counting of RBC shapes in flow usually requires an enormous effort,
and to address this issue,
there has been a proposal to apply a convolutional neural regression network to recorded microscopy images in order to automatically perform outlier-tolerant shape classification~\cite{Kihm2018}.
However, a numerical-experimental estimation of the mechanical properties of flowing cells from macro-rheological data has not been reported yet.

To address the aforementioned issues,
in this study,
we propose an integrated analysis of a model of the rheology of RBC suspensions,
coupled with macro-rheological data of the relative viscosity (non-dimensional shear viscosity) of a human RBC suspension surrounded by plasma.
In this model,
a coaxial double-cylindrical rheometer was used to obtain the shear viscosity of intact RBC and hardened RBC (HRBC) suspensions for different shear rates and volume fractions.
Along with experimental measurements,
systematic numerical analyses of RBC suspensions in shear flow were performed for a wide range of volume fractions $\phi$ (the so-called hematocrit), viscosity ratios $\lambda$ ($= \eta_1/\eta_0$), and capillary numbers $Ca = \eta_0 \dot\gamma a/G_s$,
which is the ratio between fluid viscous force and membrane elastic force.
Here, $\dot\gamma$ is the shear rate and $a$ is the major RBC radius in the resting state.
Continuum-based numerical modeling was considered,
where an isotropic hyperelastic material representing the RBC membrane is immersed in Newtonian fluids representing the cytoplasm and plasma.
Such multi-scale numerical analysis,
involving the micro to macro level,
has been verified in previous works~\cite{Takeishi2019, Takeishi2024}.
By fitting a non-linear curve of the relative viscosity of the suspensions as a function of cell volume fractions $\phi$,
the quantitative relationship between $Ca$ and $\dot\gamma$ was derived,
and then the statistical average of the membrane shear elasticities of individual intact RBCs and HRBCs was estimated as $\langle G_s \rangle =\eta_0 \dot\gamma \langle a \rangle/Ca$,
where $\langle \cdot \rangle$ denotes the average value.
The effect of viscosity ratios $\lambda$ on apparent membrane shear elasticity was also addressed.

\section{MATERIALS AND METHODS}
\subsection{\label{sec:materials}Suspension preparation}
All of the procedures were performed according to the ethical policy of Kansai University (permit number $= 21$--$71$).
Fresh human blood was sampled from two young healthy volunteers ($23$--$25$ years old) and used immediately after collection. Total three different samples ($\mathcal{N} = 3$) was prepared for each experiment.
To obtain a suspension consisting of human RBCs and plasma,
blood samples treated with anticoagulant ($3.2$\% citric sodium) were washed with phosphate-buffered saline (PBS) mixed with $1$ wt\% bovine serum albumin (BSA, Wako) and then centrifuged.
Washed RBCs were resuspended in plasma at volume concentrations (or hematocrit) of $\phi = 0.1$--$0.8$.
A whole-blood sample ($\phi = 0.4$) treated with anticoagulant (citric sodium) was also prepared for the comparison.
HRBCs were prepared by immersing normal RBCs in the aforementioned solution (PBS $+$ BSA) containing $400$ ppm ($= 0.04$\%) glutaraldehyde (GA, Sigma-Aldrich) for $30$ minutes at room temperature ($22$ $^\circ$C)~\cite{Abay2019, Kuck2022, Squier1976}.
Each measurement was performed using 3.5-ml samples.
Due to the anticoagulant,
the aggregation was not confirmed in healthy RBCs ($\phi = 0.4$ and $0.8$) suspended in plasma and whole blood before and even after the measurement with constant low shear rate of $\dot\gamma = 0.1$ s$^{-1}$ (data not shown),
where pipetting for the samples was not conducted before and after recoding microscopy images to avoid giving shear stress to the cells.

\subsection{\label{sec:exp}Experimental setup}
\begin{figure*}[t]
  \centering
  \includegraphics[height=10cm]{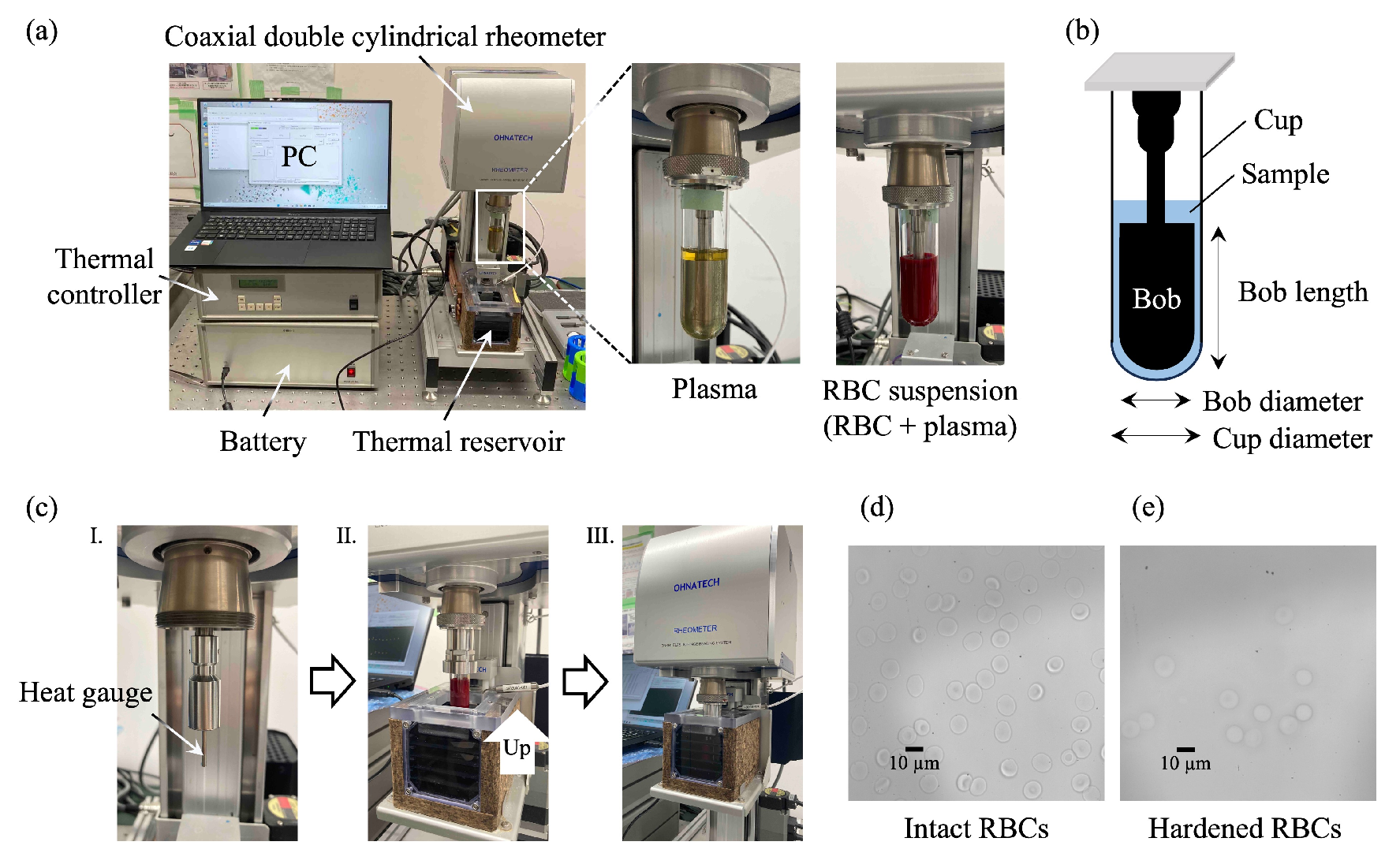}
  \caption{
  (a) Experimental setup,
  consisting of a coaxial double-cylindrical rheometer (OhnaTech Inc. Tsukuba, Ibraki, Japan), thermal reservoir, thermal controller, battery, and PC for data recording.
  Enlarged views are shown of the cup with inserted plasma ($\phi = 0$) and RBC suspension ($\phi = 0.1$), respectively.
  (b) Sketch of the bob and cup filled with a sample,
  where the diameters of the bob and cup are $19.36$ mm and $21.0$ mm, respectively,
  and the bob length, including the bottom of the hemisphere, is $103.5$ mm.
  (c) Outline of the pre-measurement procedures;
  the internal heat gauge of the bob is shown in panel (c).I;
  placing the cup into the thermal reservoir by the moving stage until top surface of the sample is completely below a volume of pure water at a room temperature of $22$ $^\circ$C (c.I\hspace{-1.2pt}I and c.I\hspace{-1.2pt}I\hspace{-1.2pt}I).
  Microscopy images of (d) intact RBCs and (e) HRBCs after measurements,
  with data obtained using a x100 lens (oil immersion).
  }
  \label{fig:setup}
\end{figure*}

The experimental apparatus consisted primarily of a coaxial double-cylindrical rheometer (ONRH-1, OhnaTech Inc. Tsukuba, Ibaraki, Japan).
The experimental setup is shown in Fig.~\ref{fig:setup}(a).
An enlarged view is shown of the plasma ($\phi = 0$) and RBC suspension ($\phi = 0.1$), both placed in a cup.
A sketch of the bob and external cylinder (cup) filled with a sample is shown in Fig.~\ref{fig:setup}(b),
where the diameters of the bob and cup are $19.36$ mm and $21.0$ mm, respectively,
and the bob length,
including the bottom of the hemisphere,
is $103.5$ mm.
The measurements were conducted after the cup was completely immersed in the thermal reservoir in pure water at a room temperature of $22$ $^\circ$C (Fig.~\ref{fig:setup}c).
During the measurements,
the sample temperatures were kept constant under the control of a system integrated with a heat gauge within the bob and thermal reservoir.
We confirmed that intact RBCs and HRBCs retained their original biconcave shapes even after measurements with a maximum shear rate of 2000 s$^{-1}$,
as shown in Figs.~\ref{fig:setup}(d) and \ref{fig:setup}(e).
Thus, we concluded that hemolysis did not affect the measurements.

\subsection{\label{sec:exp2}Rheological protocol}
At the beginning of each experiment, the samples were well mixed by pipetting.
Pre-shear was given every measurement for each shear rate with $10$--$15$ s,
which was 50\% of the total shear-duration time ($20$--$30$ s).
The time interval between the different two runs was set to be within a second.
Hence, the effect of the cell sedimentation is negligible on our experimental results not only at the beginning but also during the measurements.
Although a steady-state criterion was not defined in the measurements,
the shear-duration time in each run can be assumed as substantially enough long period to assume the steady state.
Indeed, we confirmed that the relative viscosities of RBCs suspensions agree well with previous experiments~\cite{Brooks1970, Goldsmith1972} (see Fig.~\ref{fig:mure_fitting}a).

The relevant mechanical and sample-related experimental limits of shear rheometry are plotted (e.g., low-torque limit and secondary flow effects for steady shear data, and instrument inertia effects for small-amplitude-oscillatory-shear data) in conventional bulk rheology measurements~\cite{Ewoldt2015}.
Considering these facts, aforementioned parameters regarding measurement time were empirically determined.

At each shear rate,
four runs ($n = 4$) were conducted, and thus the error bars of the relative viscosity shown in Figs.~\ref{fig:shear_viscos} and \ref{fig:validation} represent the standard deviations for total twelve measurements ($= \mathcal{N} \times n$) for each shear rate.
The measurements were well automatized, and a protocol for shear scanning in the range of $0.1$ s$^{-1}$ $\leq \dot{\gamma} \leq 10^3$ s$^{-1}$ was programed in the rheometer

\subsection{\label{sec:numerics}Flow and cell models}
We consider a cellular flow consisting of plasma, cytoplasm, and RBCs with major diameter $d$ ($= 2a = 8$ $\mu$m) and maximum thickness $2$ $\mu$m ($= a/2$) in a rectangular box with a resolution of $16$ fluid lattices per RBC radius.
The flow is assumed to be as effectively inertialess.
The RBC is modeled as a biconcave capsule, or a Newtonian fluid enclosed by a thin elastic membrane.
The RBC is initially set to have the classical biconcave shape.

The membrane is modeled as an isotropic and hyperelastic material,
and is assumed to follow the Skalak constitutive law~\cite{Skalak1973}
\begin{equation}
  \frac{w}{G_s} = \frac{1}{4} \left( I_1^2 + 2 I_1 - 2I_2 + C I_2^2\right),
  \label{SK}
\end{equation}
where $w$ is the strain energy density function,
$G_s$ is the membrane shear elastic modulus,
$C$ a coefficient representing the area incompressibility,
and $I_1$ and $I_2$ are the invariants of the strain tensor.
In the Skalak law~\eqref{SK},
the area dilation modulus is $K_s = G_s (1 + 2C)$.
In this study, we set $C = 10^2$~~\cite{Barthes-Biesel2002},
which describes an almost incompressible membrane.
Bending resistance is also considered~\cite{Li2005},
with a bending modulus $k_b = 5.0 \times 10^{-19}$ J~\cite{Puig-de-Morales-Marinkovic2007};
these values have been shown to successfully reproduce the deformation of RBCs in shear flow~\cite{Takeishi2014, Takeishi2019} and the thicknesses of cell-depleted peripheral layers in circular channels (see Figure~A.1 in~\citet{Takeishi2014}).
Although the model parameters of human healthy RBCs were reviewed in~\citet{Tomaiuolo2014},
where the bending modulus was reported as $(1.15 \pm 0.9) \times 10^{-19}$ J, we further tested different bending modulus in the rage of $O(k_b) = 10^{-20}$--$10^{-19}$ J and confirmed that the stable RBC configuration under channel flow remained the same~\cite{Takeishi2019Micromachines}.
Little effects of the variation of bending rigidity to the red cell dynamics were also reported in~\citet{Dao2006}.
Thus, we set the nonzero bending resistance $k_b$, referring to~\citet{Li2005}.

Although there was a large standard deviation of the values of the geometric and mechanical properties of healthy RBCs depending on the techniques used,
the cytoplasmic viscosity of RBCs was summarized as $6.07 \pm 3.8$ mPa~s~\cite{Tomaiuolo2014}.
The normal plasma viscosity, on the other hand, was $\eta_0 = 1.1$--$1.3$ cP ($1.1$--$1.3$ mPa~s) at 37$^\circ$C~\cite{Harkness1970}.
Hence, the physiologically relevant values of the viscosity ratio lay in the range $\lambda = \eta_1/\eta_0 = 1.89$--$8.23$ if the plasma viscosity was set to be $\eta_0 = 1.2$ cP.
The viscosity ratio $\lambda = 5$ was adopted for most of the results presented in this study, except when specifically investigating the influence of $\lambda$ on the suspension rheology.
In this case, the range $\lambda = 0.1$--$10$ was investigated.

The fluids are modeled as an incompressible Newtonian fluid that obeys the Navier--Stokes equation,
with the following governing equations of fluid velocity $\v{v}$:
\begin{align}
	\rho \left( \frac{\partial \v{v}}{\partial t} + \v{v} \cdot \nabla \v{v} \right)
	&= \nabla \cdot \mathbf{\sigma}^f  + \rho \v{f}, 
	\label{NS_eq}
	\\
	\nabla \cdot \v{v} &= 0,
	\label{continuum_eq}
\end{align}
and 
\begin{align}
\mathbf{\sigma}^f = -p\mathbf{I} + \eta \left( \nabla \v{v} + \nabla \v{v}^T \right),
\end{align}
where $\rho$ is the internal/external fluid density,
$\mathbf{\sigma}^f$ is the total stress tensor of the flow,
$\v{f}$ is the body force,
$p$ is the pressure, 
$\mathbf{I}$ is the identity tensor,
and $\eta$ is the fluid viscosity,
expressed using a volume fraction of the inner fluid $\alpha$ (0 $\leq \alpha \leq$ 1) as:
\begin{align}
        \eta = \left\{ 1 + \left( \lambda - 1 \right) \alpha \right\} \eta_0.
\end{align}
The dynamic condition requires that the load $\v{q}$ is equal to the traction jump $\left( \mathbf{\sigma}^f_\mathrm{out} - \bm{\sigma}^f_\mathrm{in} \right)$ across the membrane:
\begin{align}
	\v{q} = \left( \mathbf{\sigma}^f_\mathrm{out} - \mathbf{\sigma}^f_\mathrm{in} \right) \cdot \v{n},
\end{align}
where the subscripts ``out" and ``in", respectively, represent the outer and internal regions of the capsule,
and $\v{n}$ is the unit normal outward vector in the deformed state.

In the inertialess limit, the problem is characterized by the capillary number,
\begin{equation}
  Ca = \frac{\eta_0 \dot\gamma a}{G_s}.
  \label{Ca}
\end{equation}
To limit the computational cost and yet obtain results not affected by inertial effects,
we set the Reynolds number $Re = \rho \dot\gamma a^2/\eta_0 = 0.2$.
This value well reproduces the capsule dynamics in unbounded shear flows obtained with the boundary integral method in Stokes flow (see also Figures~12 and 13 in~\citet{Takeishi2019}).
We further checked numerical results at lower $Re$ ($= 0.05$) and confirmed that the results remain the same. (see also Figure~13 in~\citet{Takeishi2024}).

\subsection{Numerical method}
The finite-element method is used to solve the weak form of the equation governing the inertialess membrane dynamics and obtain the load $\v{q}$ acting on the membrane:
\begin{equation}
  \int_S \v{\hat{u}} \v{\cdot} \v{q} dS = \int_S \mathbf{\hat{\epsilon}} : \mathbf{T} dS,
  \label{WeakForm}
\end{equation}
where $\v{\hat{u}}$ is the virtual displacement,
$\mathbf{\hat{\epsilon}} = ( \nabla_s \v{\hat{u}} + \nabla_s \v{\hat{u}}^T )\big/2$ is the virtual strain,
and $\nabla_s (= (\mathbf{I} - \v{n}\v{n}) \cdot \nabla)$ is the surface gradient operator. 
The in-plane elastic tension $\mathbf{T}$ is obtained from the Skalak constitutive law~\eqref{SK}.
The governing equations for the fluid~\eqref{NS_eq} are discretized by the lattice-Boltzmann method (LBM) based on the D3Q19 model~\cite{Chen1998}.
The velocity at the membrane node is obtained by interpolating the velocities at the fluid node using the immersed boundary method (IBM)~\cite{Peskin2002}.
The membrane node is updated by Lagrangian tracking with the no-slip condition.
The explicit fourth-order Runge--Kutta method is used for the time integration.
In our coupling of the LBM and IBM,
the hydrodynamic interaction between individual RBCs is solved without modeling a non-hydrodynamic inter-membrane repulsive force in the case of vanishing inertia as shown in Fig.~\ref{fig:verification} in Appendix~\S\ref{appA} (see also Figure~14 in~\citet{Takeishi2024}).
The volume-of-fluid method~\cite{Yokoi2007} and front-tracking method~\cite{Unverdi1992} are employed to update the viscosity in the fluid lattices.
A volume constraint is implemented to counteract the accumulation of small errors in the volume of the individual cells~\cite{Freund2007}: in our simulation, the volume error is always kept lower than $1.0\times10^{-3}$\%.
All procedures were fully implemented on a GPU to accelerate the numerical simulation.
More precise explanations are provided in our previous works~\cite{Takeishi2019, Takeishi2022}.
The mesh size of the LBM for the fluid solution is set to be $250$ nm,
and that of the finite elements describing the membrane is also approximately $250$ nm (an unstructured mesh with $5120$ elements is used for each cell).
This resolution has been shown to successfully represent single-cell and multi-cellular dynamics~\cite{Takeishi2014}; we have verified that the results of multi-cellular dynamics do not change with twice the resolution for both the fluid and membrane~\cite{Takeishi2014}.

\subsection{Analysis}
The suspension rheology of RBCs,
or the contribution of the suspended RBCs to the bulk viscosity,
is quantified by the particle stress tensor $\bm{\Sigma}^{(p)}$~\cite{Batchelor1970}.
Specifically, for a deformable capsule and any viscosity ratio, Pozrikidis~\cite{Pozrikidis1992} analytically derived an expression for the corresponding stresslet,
so that the particle contribution to the total stress can be written as:
\begin{align}
  \bm{\Sigma}^{(p)}
  &= \frac{1}{V} \sum_{i = 1}^N \mathbf{S}_i, \\
  &= \frac{1}{V} \sum_{i = 1}^N \int_{A_i}
  \left[
  \frac{1}{2} \left( \v{r}\v{\hat{q}} + \v{\hat{q}}\v{r} \right) - \eta_0 \left( 1 - \lambda \right) \left( \v{v} \v{n} + \v{n} \v{v} \right)
  \right] dA_i,
  \label{stresslet}
\end{align}
where $V$ is the volume of the domain,
$\mathbf{S}_i$ is the stresslet of the {\it i}-th RBC (or capsule),
$\v{r}$ is the membrane position relative to the center of the RBC,
$\v{\hat{q}}$ is the load acting on the membrane (including the contribution from the bending rigidity),
$\v{v}$ is the interfacial velocity of the membrane,
and $A_i$ is the membrane surface area of the {\it i}-th RBC. 
Here, the suspension shear viscosity $\eta_\mathrm{all} (= \eta_0 + \delta \eta)$ is represented by the viscosity $\eta_0$ of the carrier fluid (plasma) and a perturbation $\delta \eta$.
This leads to the introduction of the relative viscosity $\eta_\mathrm{re}$ and the specific viscosity $\eta_\mathrm{sp}$,
which are defined as:
\begin{align}
  \eta_\mathrm{re} &= \frac{\eta_\mathrm{all}}{\eta_0} = 1 + \eta_\mathrm{sp}, \\
  \eta_\mathrm{sp} &= \frac{\delta \eta}{\eta_0} = \frac{\Sigma^{(p)}_{12}}{\eta_0 \dot\gamma_0},
  \label{mu_sp}
\end{align}
where the subscript $_1$ represents the streamwise direction and the subscript  $_2$ the wall-normal direction.

\section{Results}
\subsection{\label{sec:exp}Shear viscosity measurements}
\begin{figure*}[t]
  \centering
  \includegraphics[height=6cm]{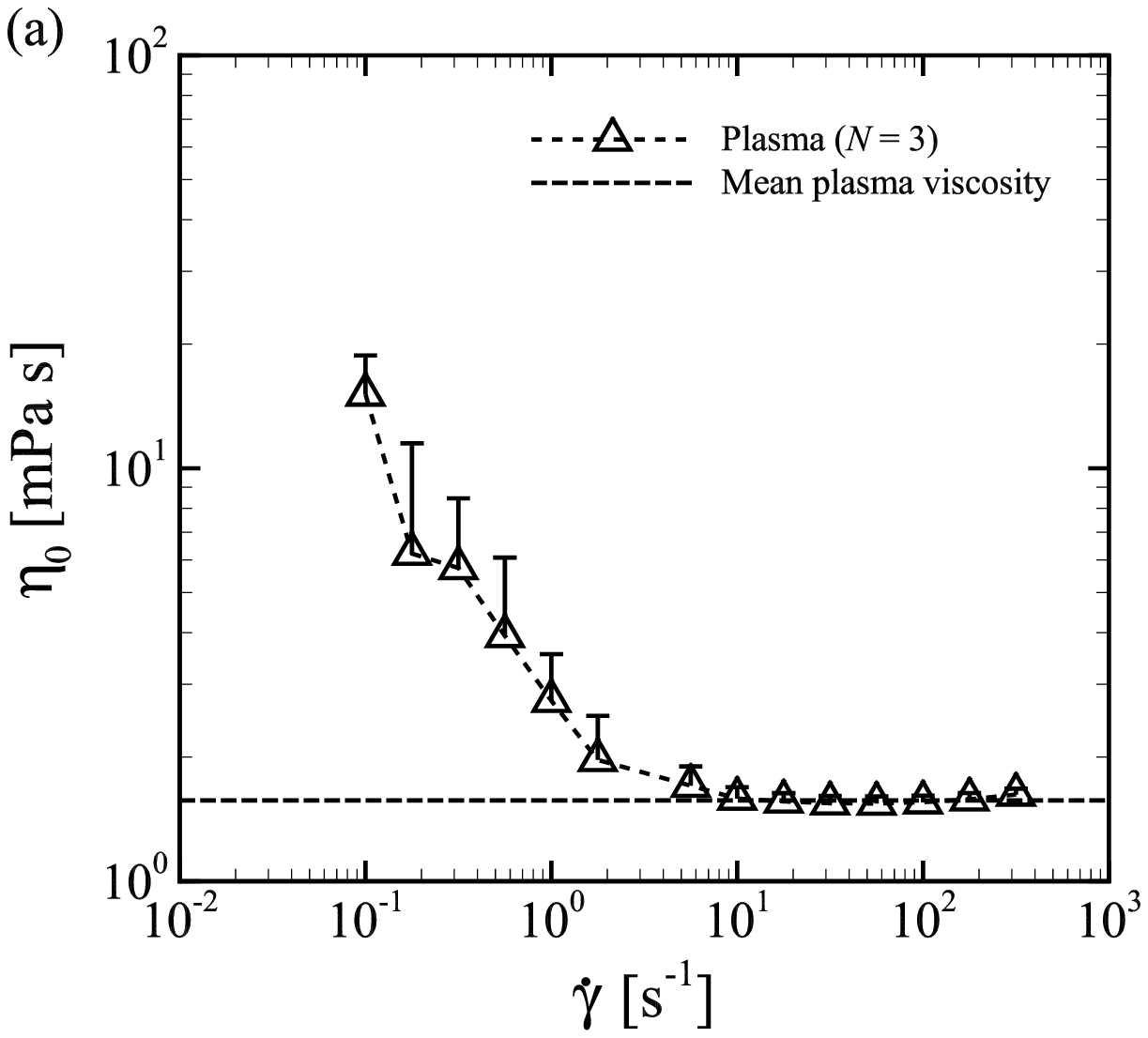}
  \includegraphics[height=6cm]{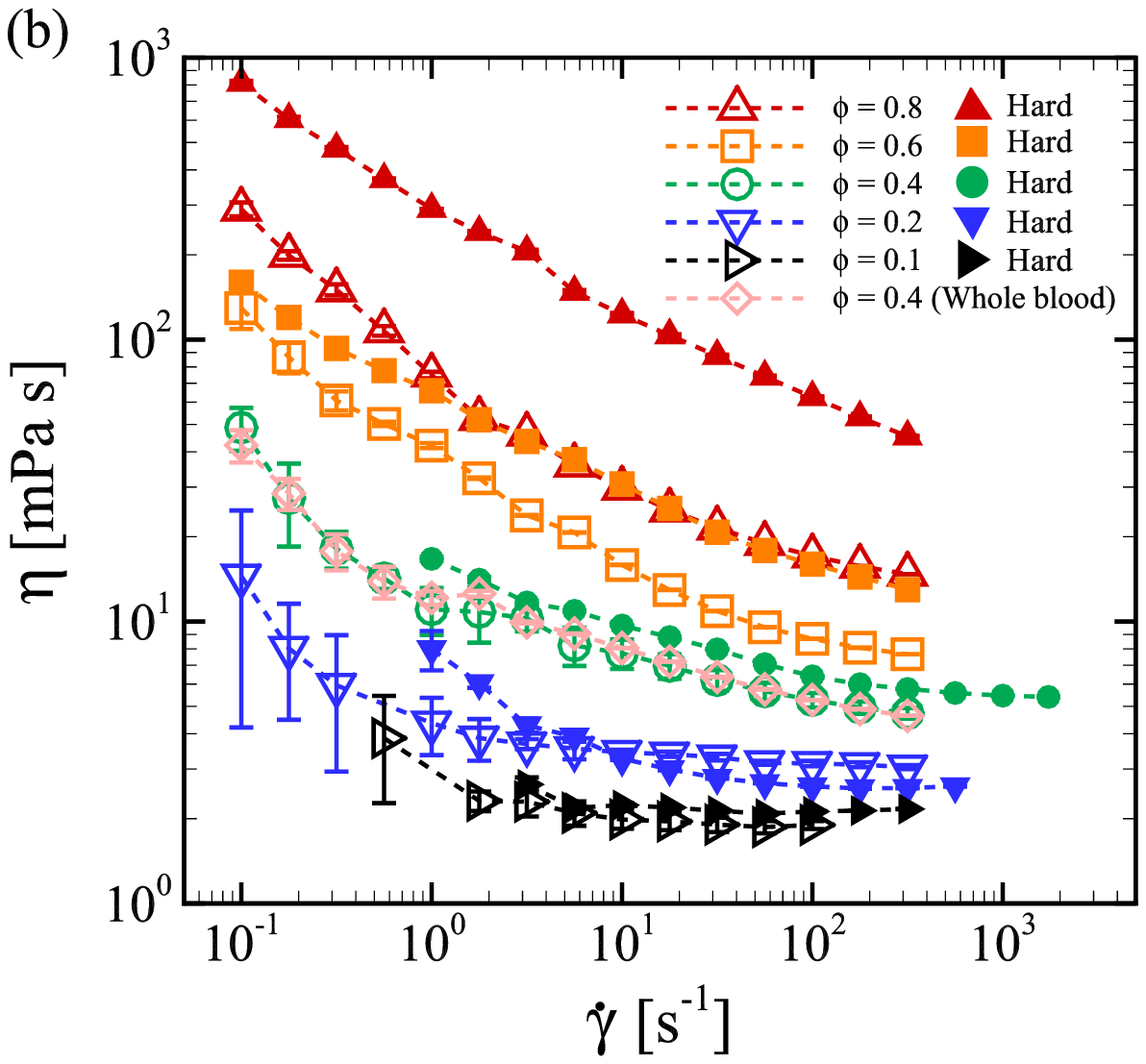}
  \caption{
  Experimental measurements of the shear viscosity of (a) the plasma $\eta_0$ and (b) intact RBC (hollow dots) and HRBC (solid dots) suspensions $\eta$ as a function of shear rates $\dot\gamma$ at $22$ $^\circ$C.
  The whole-blood samples with $\phi = 0.4$ (diamonds) are also displayed in panel (b).
  The error bars represent SDs (M $\pm$ SD., total twelve measurements, i.e., $\mathcal{N}(3) \times n(4) = 12$).
  In panel (a), SDs are displayed on only one side of the mean value to maximize clarity.
  }
  \label{fig:shear_viscos}
\end{figure*}
To determine the shear plasma viscosity $\eta_0$,
measurements were conducted with different shear rates $\dot\gamma$ ($= 0.1$--$316$ s$^{-1}$) at $22$ $^\circ$C,
and the results are shown in Fig.~\ref{fig:shear_viscos}(a),
where error bars represent the standard deviations (SDs) (M $\pm$ SD, measurements).
To increase clarity,
the error bars for $\eta_0$ are displayed on only one side of the mean value.
Although large SDs were observed at low shear rates ($\dot\gamma < 10$ s$^{-1}$),
data became stable (i.e., small SDs) at relatively high shear rates ($\dot\gamma \geq 10$ s$^{-1}$).
The shear viscosity $\eta_0$ exponentially decreased with $\dot\gamma$,
followed by a plateau for $\dot\gamma > 10$ s$^{-1}$.
Such shear-thinning behavior of human plasma was also observed in measurements using a cone-plate viscometer~\cite{Baskurt2007},
in which a film added extra drag at the radius of the cone, resulting in this apparent non-Newtonian behavior~\cite{Baskurt2007}.
Here, the film is explained as a surface layer of plasma proteins presenting at the liquid-air interface~\cite{Baskurt2007}.
Although non-Newtonian plasma characteristics were reported in previous works~\cite{Brust2013, Rodrigues2022, Varchanis2018},
we determined the mean plasma viscosity $\eta_0$ for fully saturated shear rates (i.e., $\dot\gamma > 10$ s$^{-1}$),
with reference to the aforementioned cone-plate measurements~\cite{Baskurt2007} and preliminary test described in Appendix~\S\ref{appB} (see Fig.~\ref{fig:glycerol_water}).
Then, we calculated $\eta_0$ as $\eta_0 = 1.569$ ($\pm 0.033$) mPa~s.

Experimental measurements of shear viscosities in intact RBC and HRBC suspensions are summarized in Fig.~\ref{fig:shear_viscos}(b) as a function of the shear rate $\dot\gamma$ ($0.1$ s$^{-1}$ $\leq \dot\gamma \leq 2000$ s$^{-1}$) for different volume fractions $\phi$ ($= 0.1, 0.2, 0.4, 0.6$, and $0.8$).
The results for the whole-blood samples with $\phi = 0.4$ are also plotted for comparison,
and they are almost the same as those in the intact RBC suspension (or RBCs $+$ plasma) with $\phi = 0.4$.
We thus confirmed that the experimentally observed shear-thinning behavior was caused by hydrodynamic interactions between RBCs and plasma,
i.e., that the contributions of other blood cells (e.g., white blood cells and platelets) to the shear viscosity of the blood were negligibly small.
The shear viscosity of HRBC suspensions was generally higher than those in intact RBCs.
All results exhibited the shear-thinning behavior,
which was especially significant under dense conditions ($\phi \geq 0.4$).
Under relatively small volume fractions ($\phi \leq 0.2$),
the SDs became greater at low shear rates $\dot\gamma \leq 1$ s$^{-1}$ because of technical difficulties in detecting significantly low torques.
Under dense conditions ($\phi \geq 0.4$),
however, data became stable even at low shear rates.

\subsection{\label{sim}Numerical analysis of RBC suspension}
\begin{figure*}[t]
  \centering
  \includegraphics[height=5cm]{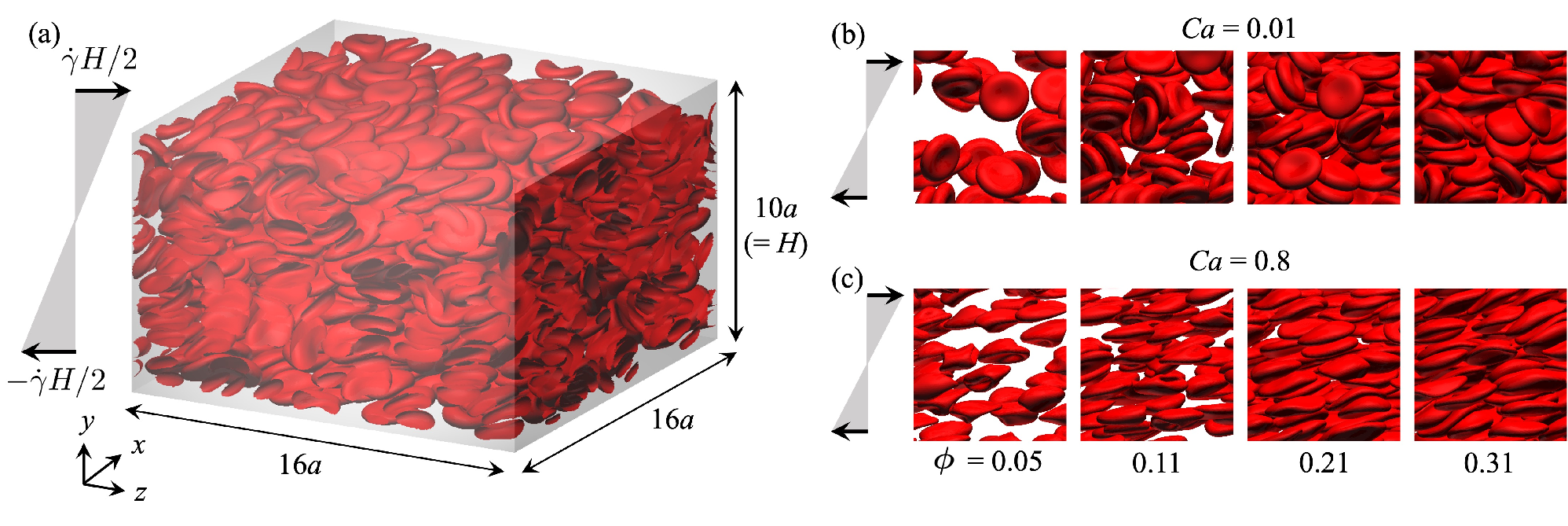}
  \includegraphics[height=4.5cm]{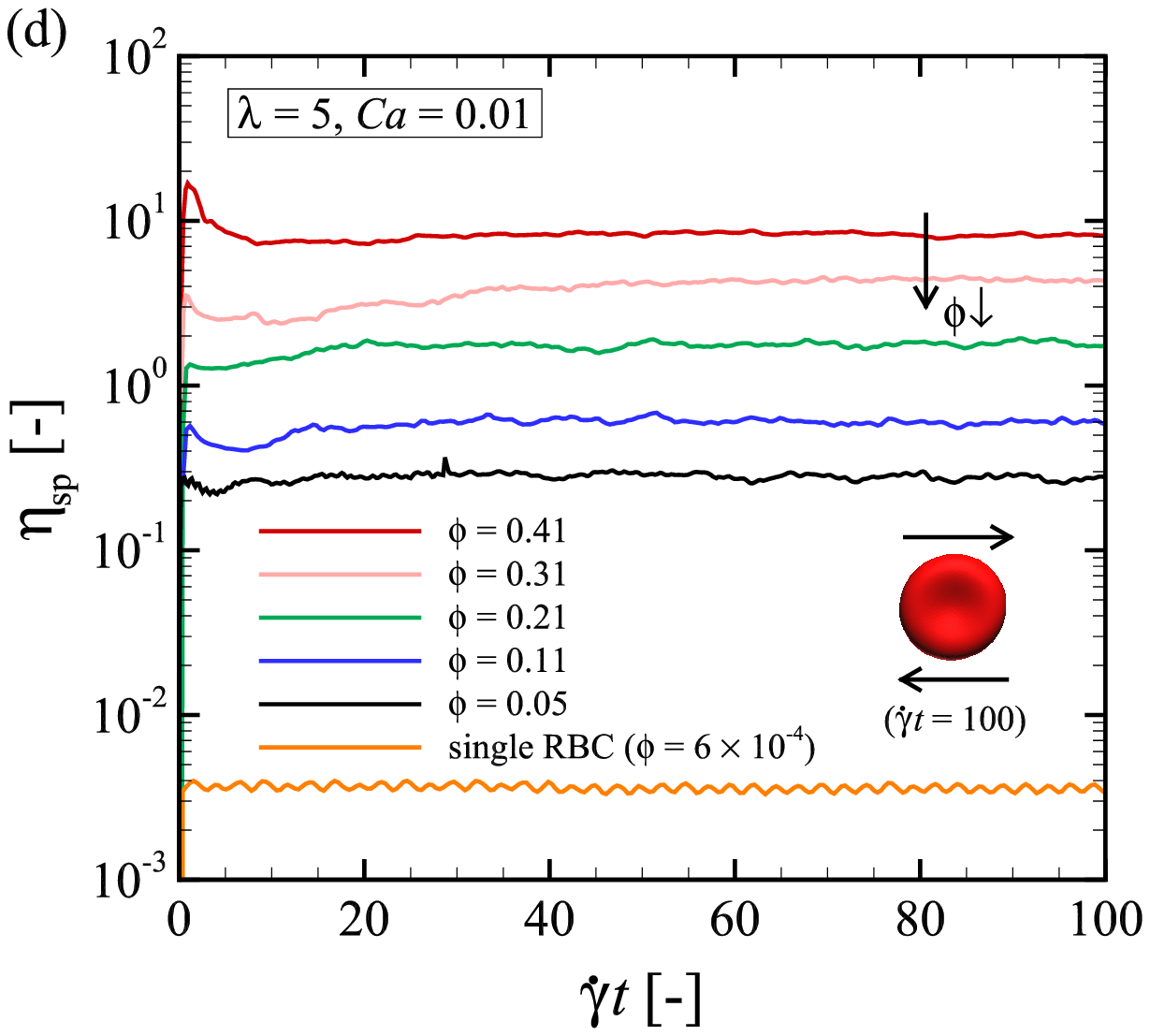}
  \includegraphics[height=4.5cm]{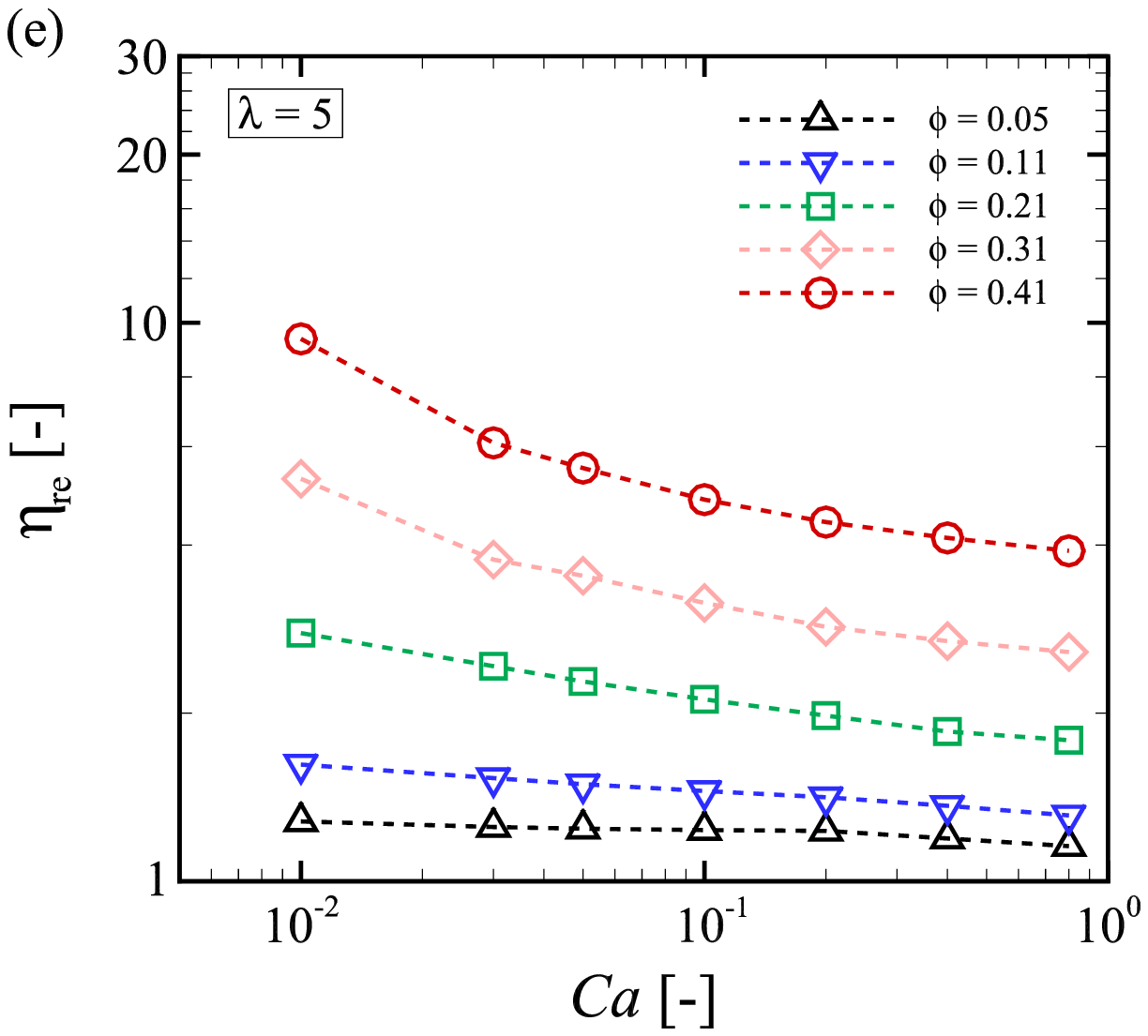}
  \includegraphics[height=4.5cm]{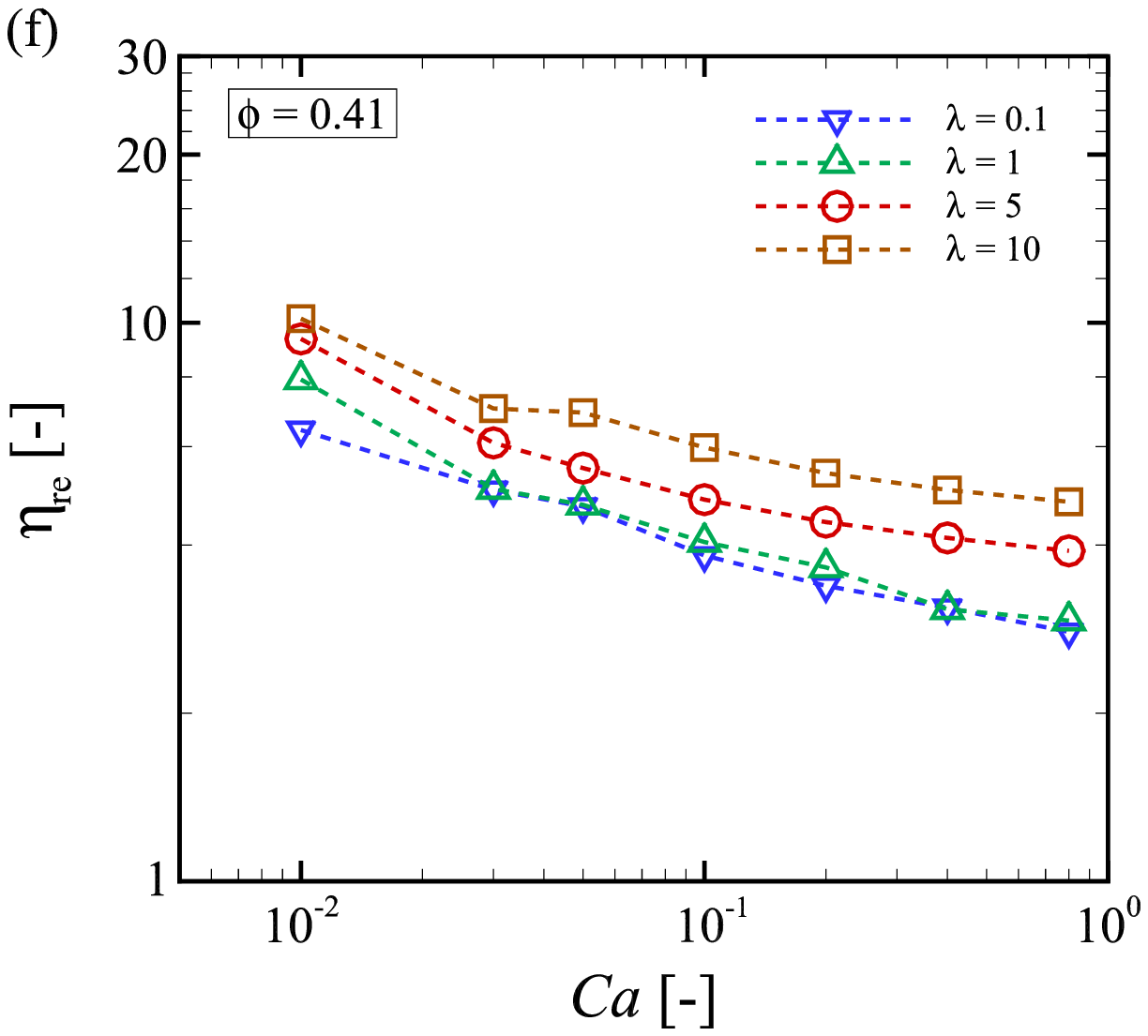}
  \caption{
  (a) Computational domain:
  a rectangular box of size $16a \times 10a \times 16a$ along the span-wise $x$, wall-normal $y$, and stream-wise $z$ directions.
  Periodic boundary conditions are imposed on the two homogeneous directions ($x$ and $z$ directions).
  The shear flow is generated by moving the top and bottom walls located at $y = \pm H/2$,
  where $H$ ($= 10a$) is the domain height. The snapshot depicts a dense suspension of RBCs ($\phi = 0.41$) with $Ca = 0.1$.
  Snapshots from our numerical results of suspensions at (b) $Ca = 0.01$ and (c) $Ca = 0.8$ with different volume fractions;
  $\phi = 0.05, 0.11, 0.21$ and $0.31$.
  (d) Time history of the specific viscosity $\eta_\mathrm{sp}$ at $Ca = 0.01$ for different $\phi$,
  where the inset represents a rolling RBC with a single stable-cell motion ($\phi = 6.135\times10^{-4}$) after sufficient time has been passed ($\dot\gamma t = 100$).
  (e) The spatiotemporal average of the relative viscosity $\eta_\mathrm{re}$ ($= 1 + \eta_\mathrm{sp}$) as a function of $Ca$ for different $\phi$.
  The results in (a)--(e) were obtained with $\lambda = 5$.
  (f) The spatiotemporal average of $\eta_\mathrm{re}$ under the dense condition $\phi = 0.41$ as a function of $Ca$ for different $\lambda$.
  }
  \label{fig:sim}
\end{figure*}

Here we consider a wall-bounded shear flow in a rectangular box of size 16$a$ $\times$ 10$a$ $\times$ 16$a$ along the span-wise $x$, wall-normal $y$, and stream-wise $z$ directions,
where $a$ is the major RBC radius.
Shear flow is generated by moving the top and bottom walls located at $y = \pm H/2$ with constant velocity $\pm \dot\gamma H/2$,
where $H$ ($= 10a$) is the domain height.
Periodic boundary conditions are imposed on the two homogeneous directions ($x$ and $z$ directions).
The size of the domain and the numerical resolution have been justified in our previous works~\cite{Takeishi2014, Takeishi2019, Takeishi2024}.
A sketch of the computational domain,
with the coordinate system used,
is shown in Fig.~\ref{fig:sim}(a) with an instantaneous visualization of one of the dense cases ($\phi = 0.41$).

Examples of snapshots of the numerical results from semi-dilute to dense conditions ($\phi = 0.05$, $0.11$, $0.21$, and $0.31$) are shown in Fig.~\ref{fig:sim}(b) for the lowest $Ca$ ($= 0.01$) and in Fig.~\ref{fig:sim}(c) for the highest $Ca$ ($= 0.8$).
Even when the viscosity ratio $\lambda$ changed,
the overall deformation shape remained the same,
and a qualitative difference was only observed at the highest $Ca$ ($= 0.8$) and under a semi-dilute condition ($\phi = 0.05$) (see Fig.~\ref{fig:snapshots} in Appendix~\S\ref{appA}),
in which the RBCs elongated with a relatively large orientation angle (defined as the angle between the flow direction and the major axis of the deformed RBC) for low $\lambda$ ($\leq 1$), while they exhibited a complex deformed shape for high $\lambda$ ($\geq 5$).
Such complex deformed shapes, so-called polylobed shapes, were also reported in~\citet{Lanotte2016}.

A single RBC subjected to low $Ca$ ($= 0.01$) shows small deformations and exhibits a rolling motion,
which is a major stable flow mode~\cite{Takeishi2019, Takeishi2022} (see inset in Fig.~\ref{fig:sim}d).
In relatively dilute suspensions ($\phi \leq 0.2$),
both the rolling and tumbling motions coexist because hydrodynamic interactions become significant before RBCs establish the rolling motion (Fig.~\ref{fig:sim}b).
In a dense suspension ($\phi = 0.41$), however, due to high packing,
the RBCs are forced to exhibit only a tank-treading (or swinging) motion,
resulting in marked elongation even for the lowest $Ca$ ($= 0.01$) (see inset in Fig.~\ref{fig:validation}a, Fig.~\ref{fig:snapshots} and Movies~S1 and S2).

The time histories of the specific viscosity $\eta_\mathrm{sp}$~\eqref{mu_sp} for different $\phi$ are shown in Fig~\ref{fig:sim}(d).
At least after non-dimensional time $\dot\gamma t \geq 40$,
the influence of the initial conditions on $\eta_\mathrm{sp}$ is negligible.
In this study, therefore, the time average starts after $\dot\gamma t \geq 40$ and continues to $\dot\gamma t \geq 100$.

The spatiotemporal average of the relative viscosity $\eta_\mathrm{re}$ as a function of $Ca$ is shown in Fig.~\ref{fig:sim}(e) at $\lambda = 5$ for different $\phi$ ($0.05 \leq \phi \leq 0.41$).
The numerical results showed shear-thinning behavior,
especially at dense conditions,
which qualitatively agrees with the experimental measurements described in Fig.~\ref{fig:shear_viscos}(b).
The results were consistent when viscosity ratios changed,
as shown in Fig.~\ref{fig:sim}(f),
where the results of $\eta_\mathrm{re}$ under the most dense condition ($\phi = 0.41$) are compared between different $\lambda$.
Although the value of $\eta_\mathrm{re}$ increased with $\lambda$ at each $Ca$,
there were no large changes for small viscosity differences $\lambda$ ($\leq 1$).

\subsection{\label{mure_phi}Estimation of membrane shear elasticity from rheological data}
\begin{figure*}[t]
  \centering
  \includegraphics[height=6cm]{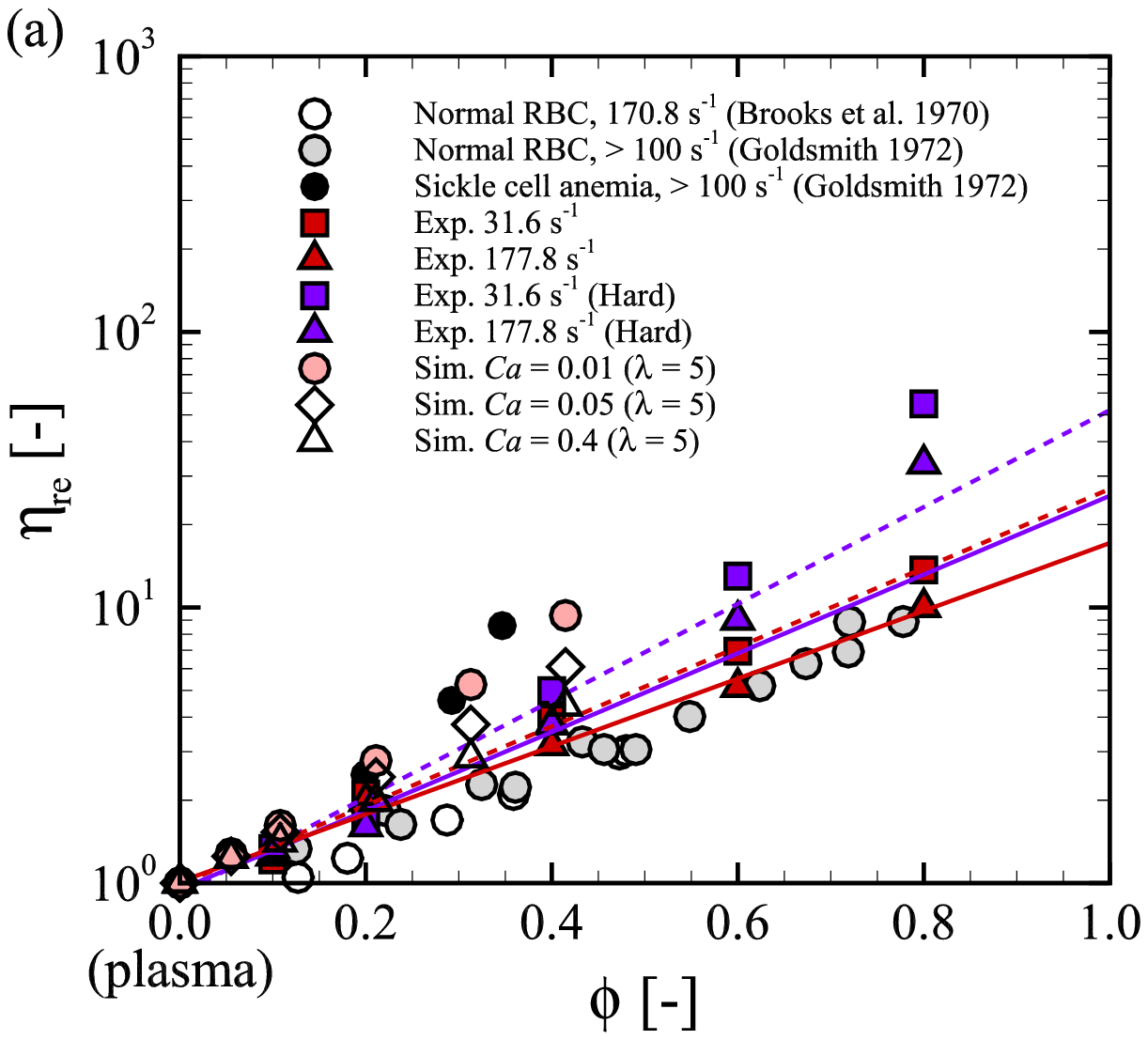}
  \includegraphics[height=6cm]{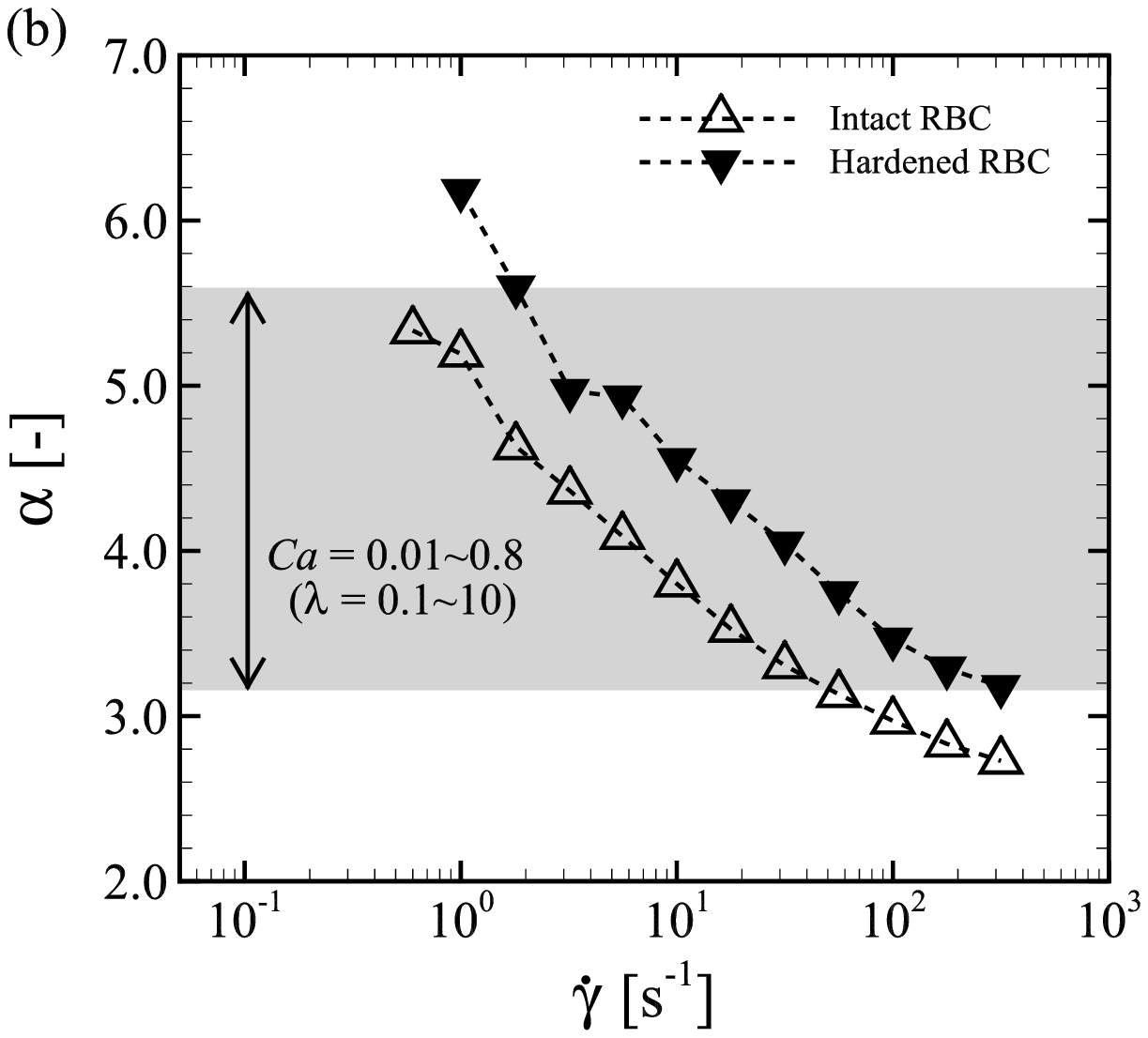}
  \includegraphics[height=6cm]{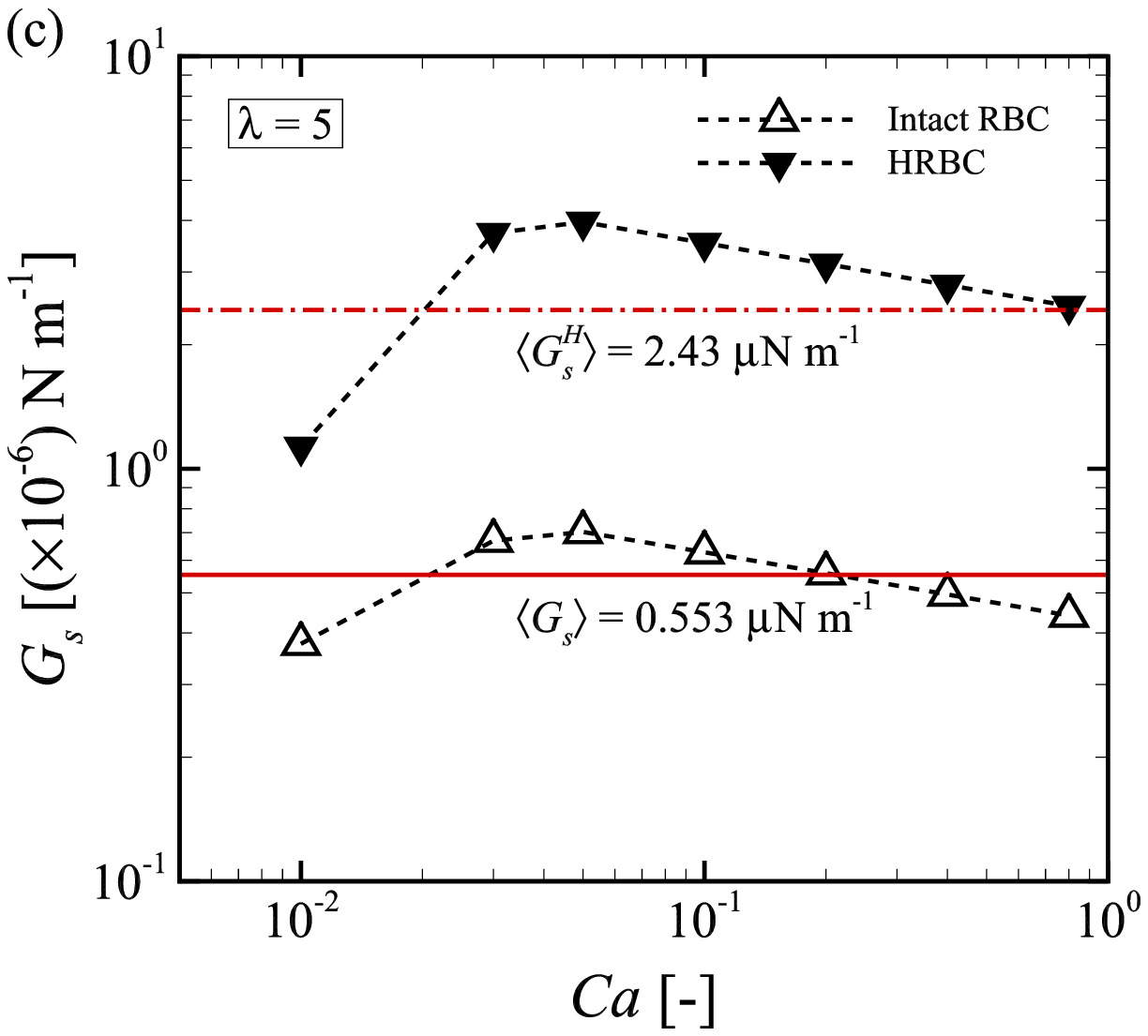}
  \caption{
  (a) The relative viscosity $\eta_\mathrm{re}$ of suspensions of intact RBCs and HRBCs as a function of volume fraction $\phi$ for different  shear rates $\dot\gamma$.
  Our experimental and numerical results are displayed,
  along with the previous data of normal human RBCs in plasma for $\dot\gamma = 170.8$ s$^{-1}$~~\cite{Brooks1970, Goldsmith1972},
and data of normal/sickle human RBC suspension for $\dot\gamma > 100$ s$^{-1}$~~\cite{Goldsmith1972}.
  Lines are the approximation curves given as $\eta_\mathrm{re} = \exp{(\alpha \phi)}$ (dashed lines for $\dot\gamma = 31.6$ s$^{-1}$ and solid lines for $\dot\gamma = 177.8$ s$^{-1}$),
  where the coefficient $\alpha$ is given by least-square fitting to the plot of $[\eta_\mathrm{re} - \exp{(\alpha \phi)}]$.
  (b) The coefficients $\alpha$ in the approximation curves $\eta_\mathrm{re} = \exp{(\alpha \phi)}$ for intact RBC and HRBC suspensions are shown as a function of the shear rate $\dot\gamma$.
  The gray region denotes the range of $\alpha$ obtained with numerical simulations for different $Ca$ ($= 0.01$--$0.8$) and different $\lambda$ ($= 0.1$--$10$).
  (c) Estimated membrane shear elasticity $G_s$ at each $\lambda = 5$ as a function of $Ca$. The averaged $G_s$ in each suspension is shown as a solid line for intact RBCs ($\langle G_s \rangle = 0.553$ $\mu$N~m$^{-1}$) and as a dashed-dot line for HRBCs ($\langle G_s^H  \rangle = 2.43$ $\mu$N~m$^{-1}$).
  }
  \label{fig:mure_fitting}
\end{figure*}

The relative viscosity $\eta_\mathrm{re}$ was calculated from the shear viscosity shown in Fig.~\ref{fig:shear_viscos}(b) and normalized by the plasma viscosity,
and then replotted in Fig.~\ref{fig:mure_fitting}(a) as a function of $\phi$,
where $\eta_\mathrm{re} = 1$ at $\phi = 0$ (no RBCs), i.e., only plasma.
The numerical results obtained with $\lambda = 5$ for specific cases of $Ca$ are also displayed,
as well as the experimental results for a suspension of normal human RBCs in plasma for $\dot\gamma = 170.8$ s$^{-1}$~~\cite{Brooks1970, Goldsmith1972},
and for a suspension of normal/sickle human RBCs for $\dot\gamma > 100$ s$^{-1}$~~\cite{Goldsmith1972}.
Our experimental measurements,
in particular at $\dot\gamma = 177.8$ s$^{-1}$,
agreed well with previous works~\cite{Brooks1970, Goldsmith1972},
and also agreed well with the numerical results obtained at $Ca = 0.4$.
The relative viscosity of the sickle RBC suspension~\cite{Goldsmith1972},
which is the largest values of all plots,
agreed with the numerical results at the lowest $Ca$ ($= 0.01$).

Approximation curves $\eta_\mathrm{re} = \exp{(\alpha \phi)}$ were superposed on the aforementioned plots of our experimental measurements,
and the coefficient $\alpha$ was given by least-square fitting to the plot of $[\eta_\mathrm{re} - \exp{(\alpha \phi)}]$.
Note that for all data, the coefficient of each determination was $R^2 > 0.9$.
For intact RBC suspensions, all data $\phi$ ($\leq 0.8$) were used for the fitting,
while for HRBC suspensions, data of limited $\phi$ ($\leq 0.4$) were used because the fit was not good under such dense conditions $\phi > 0.4$.
Indeed, plots of HRBCs for $\phi \geq 0.6$ deviated from the approximation curves (Fig.~\ref{fig:mure_fitting}a), as also seen in a previous study~\cite{Goldsmith1972}.
This result may suggest that a weak jamming transition occurs in HRBC suspensions for $\phi > 0.4$.

The coefficients $\alpha$ in the approximation curves in Fig.~\ref{fig:mure_fitting}(a) are summarized in Fig.~\ref{fig:mure_fitting}(b) as a function of the shear rate $\dot\gamma$.
The values of $\alpha$ decreased with $\dot\gamma$,
and the results reconfirm the shear-thinning behavior in suspensions of both intact RBCs and HRBCs.
The coefficients $\alpha$ obtained with numerical simulations for different $Ca$ were in the range of $0.6$ s$^{-1}$ $\leq \dot\gamma \leq$ $316.3$ s$^{-1}$ for all $\lambda$ ($= 0.1$--$10$).
Comparing $\alpha$ between experimental and numerical results,
we analyzed the relationship between $Ca$ and the shear rate $\dot\gamma$ for each intact RBC and HRBC suspension.
For instance,
$Ca = 0.4$ corresponded to $\dot\gamma = 31.6$ s$^{-1}$ in intact RBCs,
and to $\dot\gamma = 177.8$ s$^{-1}$ in HRBCs.
The estimated value of $G_s$ at each $Ca$ for intact RBCs and HRBCs is shown in Fig.~\ref{fig:mure_fitting}(c),
where the major RBC radius $\langle a \rangle = 4$ $\mu$m was used.
Although it is ideal to have a unique estimated value of $G_s$ at each $Ca$ ($= 0.01$--$0.8$),
these values may possess certain errors, e.g., due to the number of fitting data and the relatively large errors at lower shear rates $\dot\gamma$.
Using seven $G_s$ values for each $Ca$ ($= 0.01$--$0.8$) in each suspension,
we derived the mean membrane shear elasticity as $\langle G_s \rangle = \eta_0 \dot\gamma \langle a \rangle/Ca = 0.553$ $\mu$N~m$^{-1}$ for intact RBCs,
and $\langle G_s^H \rangle = 2.43$ $\mu$N~m$^{-1}$ for HRBCs.
We found that the two estimated $G_s$ values differed by one order of magnitude.

\subsection{Validation of estimated membrane shear elasticity}
\begin{figure*}[t]
  \centering
  \includegraphics[height=6cm]{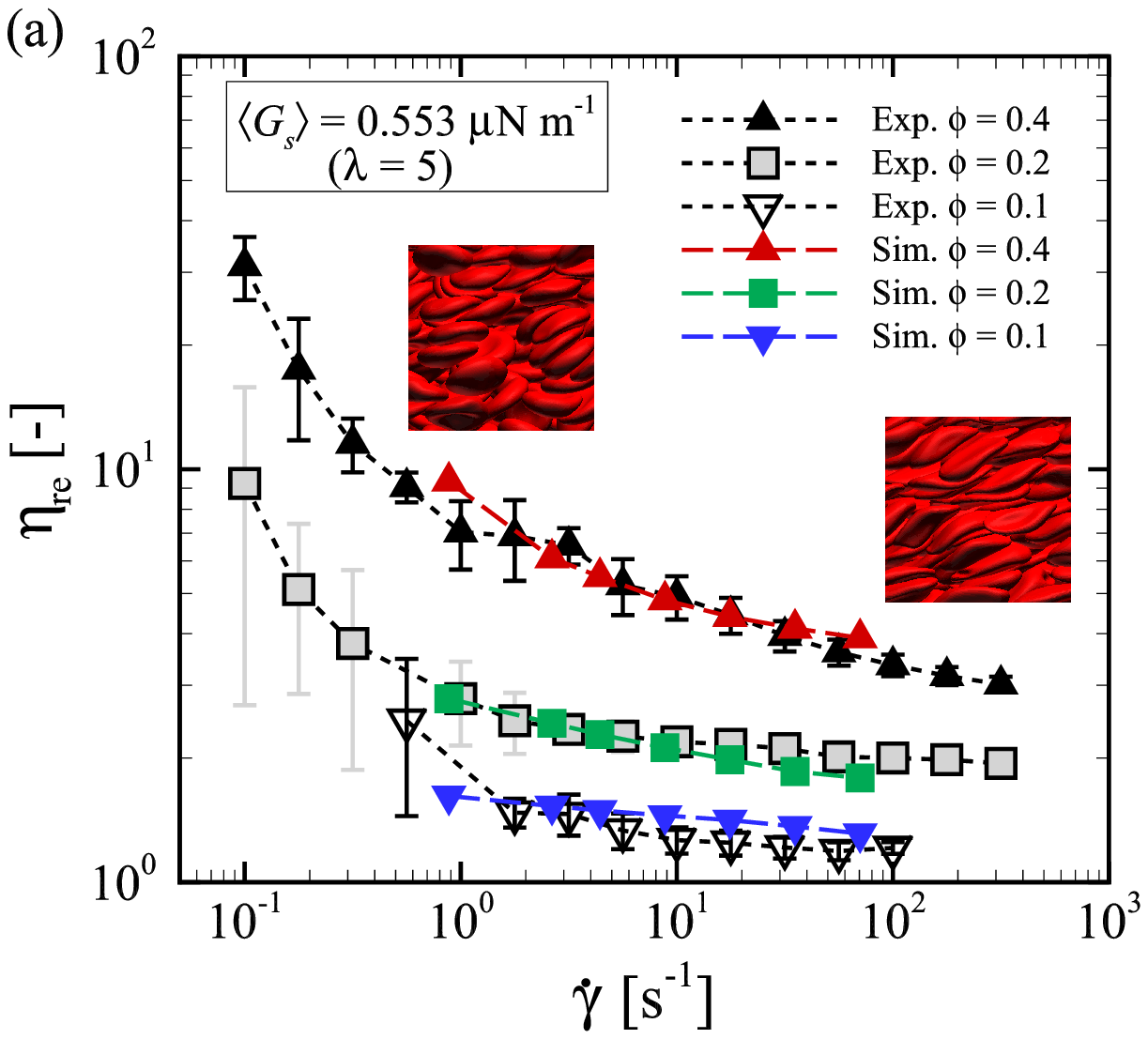}
  \includegraphics[height=6cm]{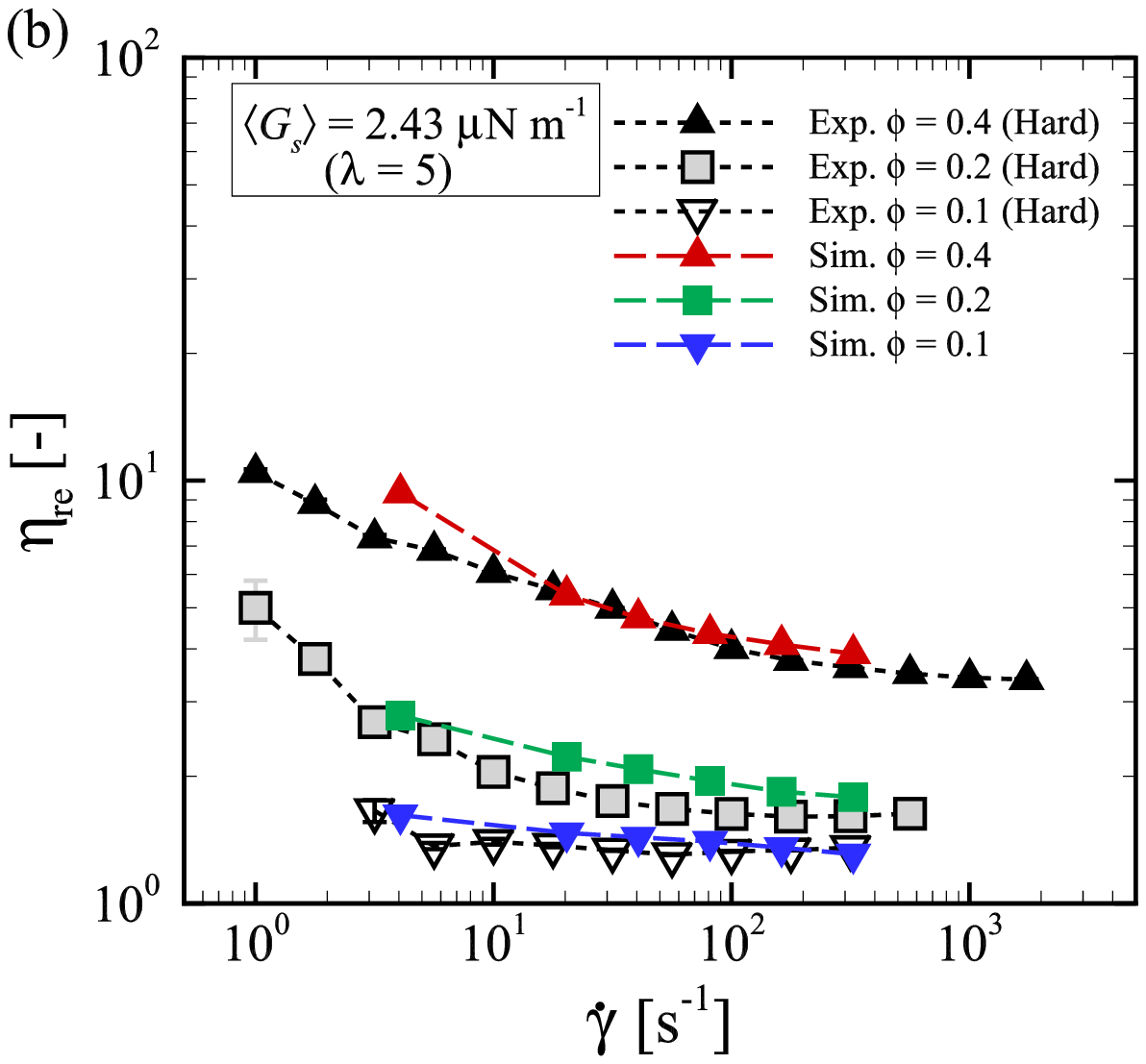}
  \includegraphics[height=6cm]{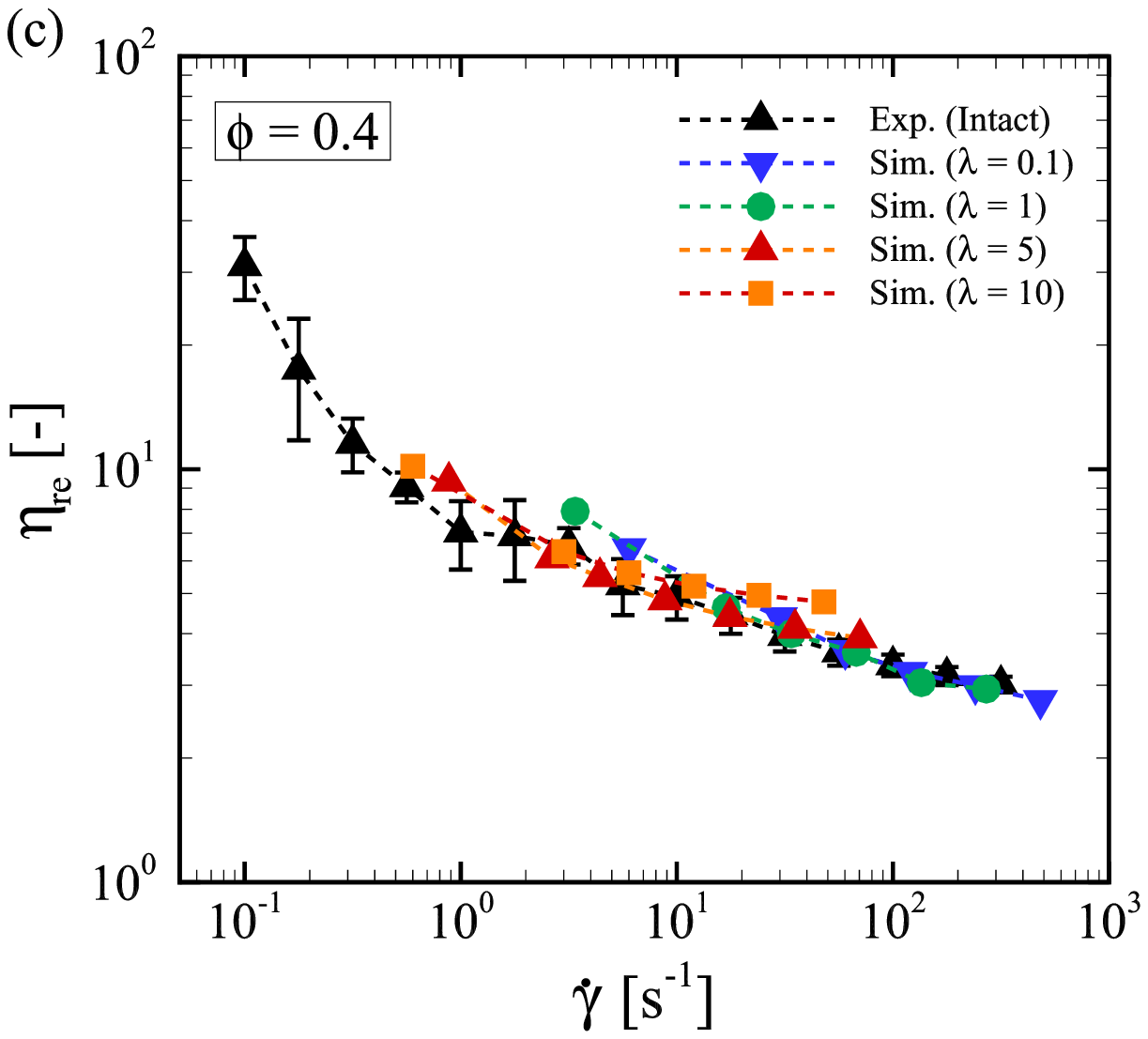}
  \includegraphics[height=6cm]{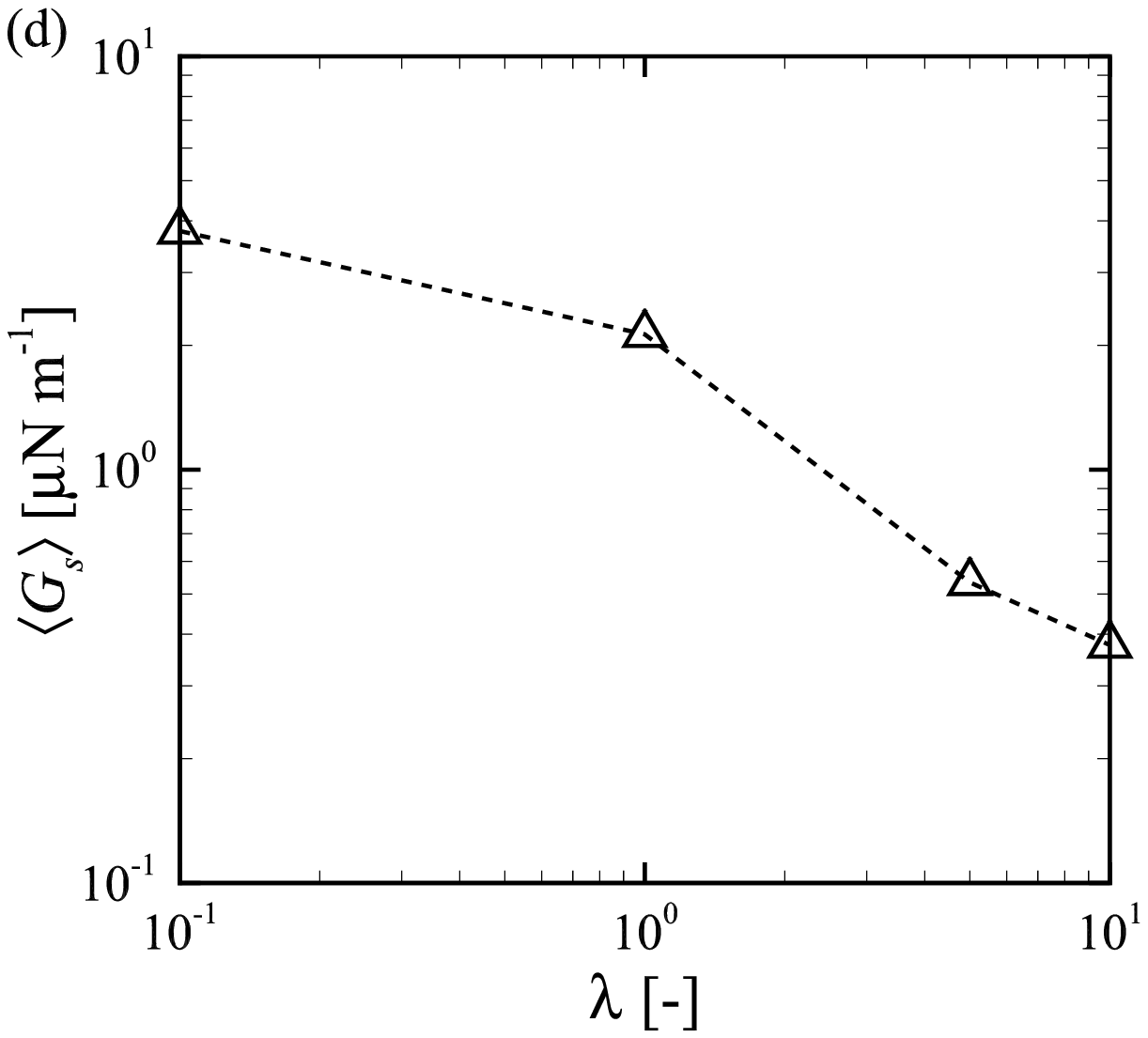}
  \caption{
  Comparison of the relative viscosity $\eta_\mathrm{re}$ between (a) intact RBC suspensions and (b) HRBC suspensions and numerical results with $\lambda = 5$ as a function of the shear rate $\dot\gamma$ for different volume fractions $\phi$.
  Each shear rate corresponding to a specific $Ca$ was calculated using $\langle G_s \rangle$ ($= 0.553$ $\mu$N~m$^{-1}$) for intact RBCs and $\langle G_s^H \rangle$ ($= 2.43$ $\mu$N~m$^{-1}$) for HRBCs.
  The insets in panel (a) are representative snapshots for $\phi = 0.4$ at the lowest and highest $Ca$ ($= 0.01$ and $0.8$, respectively) (see also Movies S1 and S2 for each $Ca$).
  (c) Comparison of the relative viscosity $\eta_\mathrm{re}$ between intact RBC suspensions and numerical simulations for different $\lambda$ as a function of the shear rate $\dot\gamma$ under the dense condition ($\phi = 0.4$).
  (d) Estimated membrane shear elasticity $\langle G_s \rangle$ as a function of $\lambda$.
  }
  \label{fig:validation}
\end{figure*}
Since the aforementioned $\langle G_s \rangle$ (or $\langle G_s^H \rangle$) values are based on a wide range of cell volume fractions $\phi$ spanning from relatively dilute $\phi$ ($= 0.1$) to dense $\phi$ ($= 0.8$) conditions, 
it is expected that these estimated values potentially reproduce the experimental results of the relative viscosity $\eta_\mathrm{re}$,
at least for the measurable range of $\phi$.
The estimated $\langle G_s \rangle$ is now validated by comparing $\eta_\mathrm{re}$ between experimental and numerical results.
Figure~\ref{fig:validation}(a) shows the comparison of $\eta_\mathrm{re}$ obtained with experimental measurements for intact RBC suspensions and numerical simulations for $\lambda = 5$.
Our numerical results for $\eta_\mathrm{re}$ agreed well with the experimental measurements independently of $\phi$,
and each shear rate was related to the specific $Ca$ by $\langle G_s \rangle$ ($= 0.553$ $\mu$N~m$^{-1}$).
The insets represent enlarged views of dense suspensions ($\phi = 0.4$) at the lowest and highest $Ca$ ($= 0.01$ and $0.8$, respectively).
Furthermore, the estimated $\langle G_s^H \rangle$ ($= 2.43$ $\mu$N~m$^{-1}$) for HRBCs could also qualitatively reproduce experimental measurements,
as shown in Fig.~\ref{fig:validation}(b).
These results suggest that the statistical average of the membrane shear elasticity of both intact and hardened RBCs can be estimated from macro-rheological data.

To address the question of {\it how the viscosity ratio $\lambda$ between the cytoplasm and plasma alters the relative membrane shear elasticity at the single-cell level},
we performed numerical simulations for different viscosity ratios $\lambda$ ($= 0.1, 1$, and $10$) and different volume fractions $\phi$.
Following the same aforementioned strategy, 
we estimated $\langle G_s \rangle$ for these series of numerical results,
and compared $\eta_\mathrm{re}$ in intact RBC suspensions at $\phi = 0.4$,
as shown in Fig.~\ref{fig:validation}(c).
Overall, the numerical results of $\eta_\mathrm{re}$ agreed well with the experimental measurements independently of $\lambda$.
The fit was good with relatively higher shear ranges ($\dot\gamma > 10$ s$^{-1}$) for small $\lambda$ ($\leq 1$),
while was good with low shear ranges ($\dot\gamma < 10$ s$^{-1}$) for large $\lambda$ ($\geq 5$).
The fit with $\lambda = 5$ exhibited the best agreement with the experimental results for a broad range of shear rates.
This is consistent with experiments determining that the physiologically relevant viscosity ratio was $1.89$--$8.23$ ($\eta_1 = 6.07 \pm 3.8$ mPa~s~\cite{Tomaiuolo2014} and $\eta_0 = 1.2$ mPa~s~\cite{Harkness1970} were assumed here).
The estimated $\langle G_s \rangle$ decreased with $\lambda$,
as shown in Fig.~\ref{fig:validation}(d).
The changes in $\langle G_s \rangle$ were within one order of magnitude,
at least for $0.1 \leq \lambda \leq 10$.

\section{Discussion and conclusions}
To utilize the shear viscosity of human RBC suspensions to estimate individual RBC deformability,
as represented by membrane shear elasticity $G_s$,
we performed an analysis of an integrated model of the rheology of human RBC suspensions.
In this model,
the relative viscosities $\eta_\mathrm{re}$ of intact or hardened RBC suspensions obtained using a coaxial double-cylindrical rheometer were compared with numerical results.
The estimated membrane shear elasticities of individual intact RBCs reproduced well the experimentally observed shear-thinning behavior in the suspensions.
Using limited volume fractions without cell jamming,
the approach was still applicable to HRBC suspensions.
Therefore, we conclude that the statistical average of individual RBC deformability can be estimated from macro-rheological data.

Membrane shear elasticity $G_s$ can usually be estimated by a stretch test of a single RBC under steady state~\cite{Suresh2005}.
Through the procedure,
we calculated $G_s$ as $G_s = 4$ $\mu$N~m$^{-1}$ for intact RBCs~\cite{Takeishi2014, Takeishi2019}.
An estimation using macro-rheological data in this study,
however, showed that  $\langle G_s \rangle$ for intact RBCs was one order of magnitude smaller ($O(G_s) = 10^{-7}$ N~m$^{-1}$) than that obtained by the stretch test.
The result may suggest that conventional stretch tests potentially overestimate the membrane shear elasticity $G_s$.
Although it is expected that our numerical-experimental estimation of $G_s$ based on macro-rheological data inevitably differ to some degree,
these results prove the possibility of retrograde estimation from the macro to micro properties of cells.

For HRBCs treated with $400$ pm GA,
the value of $\langle G_s^H \rangle$ was almost four times greater than in the intact RBCs,
with $\langle G_s^H \rangle/\langle G_s \rangle = 4.39$.
In this study,
the estimated $\langle G_s \rangle$ was obtained by averaging seven cases, i.e., seven different $Ca$ conditions (Fig.~\ref{fig:mure_fitting}c).
Although its order of magnitude was not changed by the averaging of different cases,
the accurate estimation of $G_s$ requires separate data samples for different $Ca$.

Comparisons with experimental and numerical results of relative viscosity $\eta_\mathrm{re}$ were performed for $O(\dot\gamma) = 10^{-1}$ s$^{-1}$--$10^2$ s$^{-1}$.
Highly accurate predictions for much low shear rates $\dot\gamma \leq 10$ s$^{-1}$,
which potentially induces RBC aggregations, may require not only an aggregation model such as in~\citet{Fedosov2011} but also overcoming the sensitivity limits of the rheometer,
which are however our future study.
A recent numerical simulation quantified the dynamics of RBC aggregations under shear flow~\cite{Abbasi2021}.
Thus, it will be interesting to study whether the order of magnitude of aggregation forces can be estimated from shear-thinning behavior at low shear rates obtained with experimental measurements.

As mentioned in Fig.~\ref{fig:mure_fitting}(a),
the fit works only for $\phi \leq 0.4$ in the case of HRBCs.
Our numerical-experimental estimation of $\langle G_s \rangle$ is limited in cases with no jamming effect.
Due to numerical limitations,
we could not perform an analysis of relatively dense cases ($\phi > 0.41$).
Thus, it remains an open question whether the deformability of individual RBCs can be estimated from rheological data under massively dense conditions that can potentially cause jamming transitions.
The next challenge may be to extend this methodology to jammed conditions, and to hardened cell suspensions.

Since it is known that GA exposure causes cross-linking of RBC proteins~\cite{Abay2019, Kuck2022, Squier1976},
a certain increase of $\langle G_s \rangle$ is a hallmark of reduced cell membrane flexibility caused by the cross-linking of membrane and intracellular proteins.
Considering shear-induced ATP release from RBCs~\cite{Eguchi1997, Forsyth2011},
our achievements may clarify the levels of intracellular compounds that are known to relate to organ-scale metabolism~\cite{Teruya2021}.
This study provides the first conclusive evidence that the statistical average of the mechanical properties of individual RBCs can be estimated from macro-rheological data.
This knowledge will be helpful for formulating future medical diagnoses at the single-cell level using mechanics-based AI.

\section*{Supplementary material}
\noindent
Movie\_S1(.mov) : RBCs in wall-bounded shear flow at $\phi = 0.41$, $\lambda = 5$ and $Ca = 0.01$.
Movie\_S2.(mov) : RBCs in wall-bounded shear flow at $\phi = 0.41$, $\lambda = 5$ and $Ca = 0.8$.

\begin{acknowledgments}
This research was supported by JSPS KAKENHI Grant Numbers JP20H02072, and JSPS24K00809.
This study was partially funded by Daicel Corporation.
Finally, N.T thanks Dr. Ryoko Otomo for helpful discussions and technical supports.
\end{acknowledgments}

\section*{AUTHOR DECLARATIONS}
\subsection*{Conflict of Interest}

The authors report no conflict of interest.

\subsection*{Author Contributions}
\textbf{Naoki Takeishi:} Conceptualization (lead); Funding acquisition (lead); Data curation (lead); Formal analysis (lead); Investigation (lead); Methodology (lead); Resources (lead); Software (lead); Validation (lead); Visualization (lead); Writing - original draft (lead); Writing - review \& editing (equal).
\textbf{Kodai Ngaishi, Tomohiro Nishiyama:} Data curation (equal); Formal analysis (equal); Investigation (equal); Methodology (equal); Validation (equal); Writing - review \& editing (supporting).
\textbf{Takeshi Nashima:} Investigation (supporting); Methodology (supporting); Resources (lead); Software (lead); Writing - review \& editing (supporting).
\textbf{Masako Sugihara-Seki:} Conceptualization (supporting); Funding acquisition (lead); Data curation (supporting); Formal analysis (supporting); Investigation (supporting); Methodology (supporting); Validation (supporting); Writing - original draft (supporting); Writing - review \& editing (equal).

\appendix

\section{\label{appA}Numerical verification and simulation results}
\begin{figure}
  \centering
  \includegraphics[height=6cm]{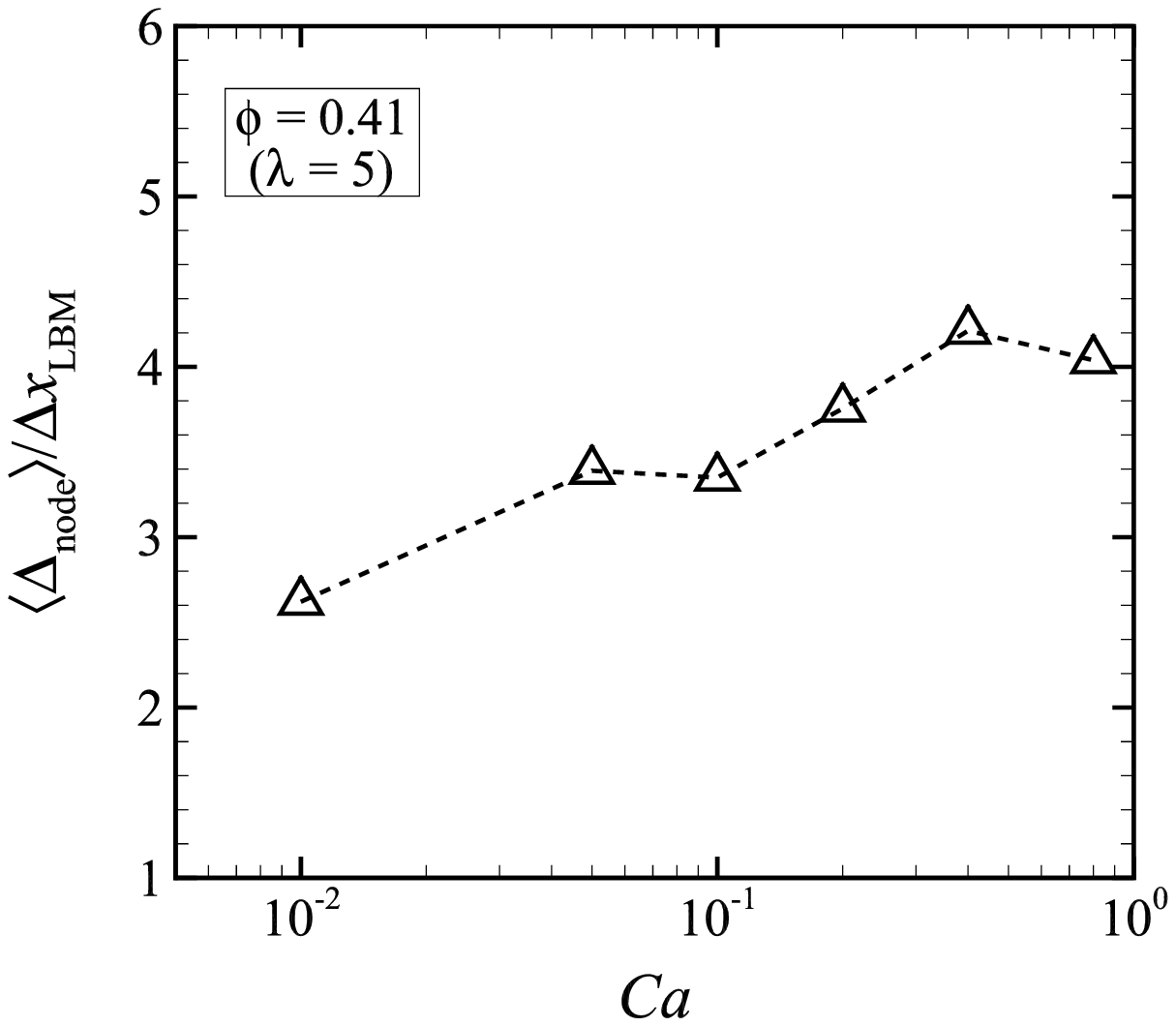}
  \caption{
  The space-temporal average of the distance between neighboring membrane nodes $\langle \Delta_\mathrm{node} \rangle/\Delta x_\mathrm{LBM}$ as a function of $Ca$,
  where $\Delta x_\mathrm{LBM}$ is the mesh size of the LBM.
  The results are obtained with volume fraction $\phi = 0.41$ and viscosity ratio $\lambda = 5$.
  }
  \label{fig:verification}
\end{figure}

We investigated the distance between neighboring membrane nodes.
The result of spatiotemporal average of the distance of neighboring membrane nodes $\langle \Delta_\mathrm{node} \rangle$ for different $Ca$ is shown in Fig~\ref{fig:verification}.
The result is obtained with $\phi = 0.41$, and $\lambda = 5$.
Even at the smallest $Ca$ ($= 0.01$), namely, the hardest RBC, the average distance still maintains over two fluid lattices. 

\begin{figure}
  \centering
  \includegraphics[height=8cm]{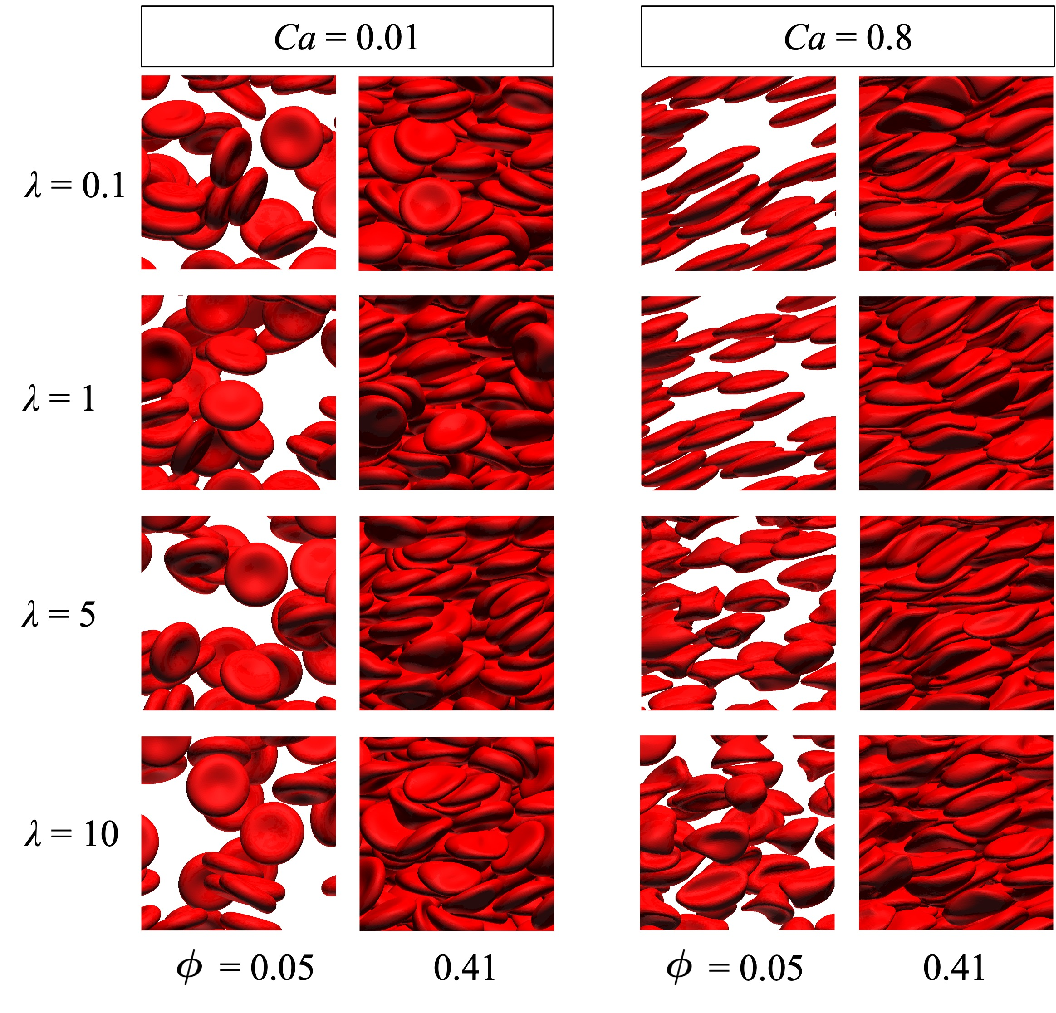}
  \caption{
  Snapshots from our numerical results of suspensions with different viscosity ratios $\lambda$ ($=  0.1$, $1$, $5$, and $10$) and different volume fractions $\phi$ ($= 0.05$ and $0.41$) at the imposed lowest $Ca$ ($= 0.01$, left panels) and highest $Ca$ ($= 0.8$, right panels).
  }
  \label{fig:snapshots}
\end{figure}

Individual RBCs from semi-dilute ($\phi = 0.05$) to dense ($\phi = 0.41$) suspensions are simulated for different $Ca$,
and examples of snapshots of the numerical results are shown in figure~\ref{fig:snapshots}.
The RBCs elongated with a relatively large orientation angle (defined as the angle between the flow direction and the major axis of the deformed RBC) for low viscosity ratios $\lambda$ ($\leq 1$),
while they exhibited a complex deformed shape for high viscosity ratios $\lambda$ ($\geq 5$).
Such complex deformed shapes, so-called polylobed shapes, were also reported in~\cite{Lanotte2016}.

\section{\label{appB} Sensitivity limits of the rheometer}
\begin{figure}
  \centering
  \includegraphics[height=6cm]{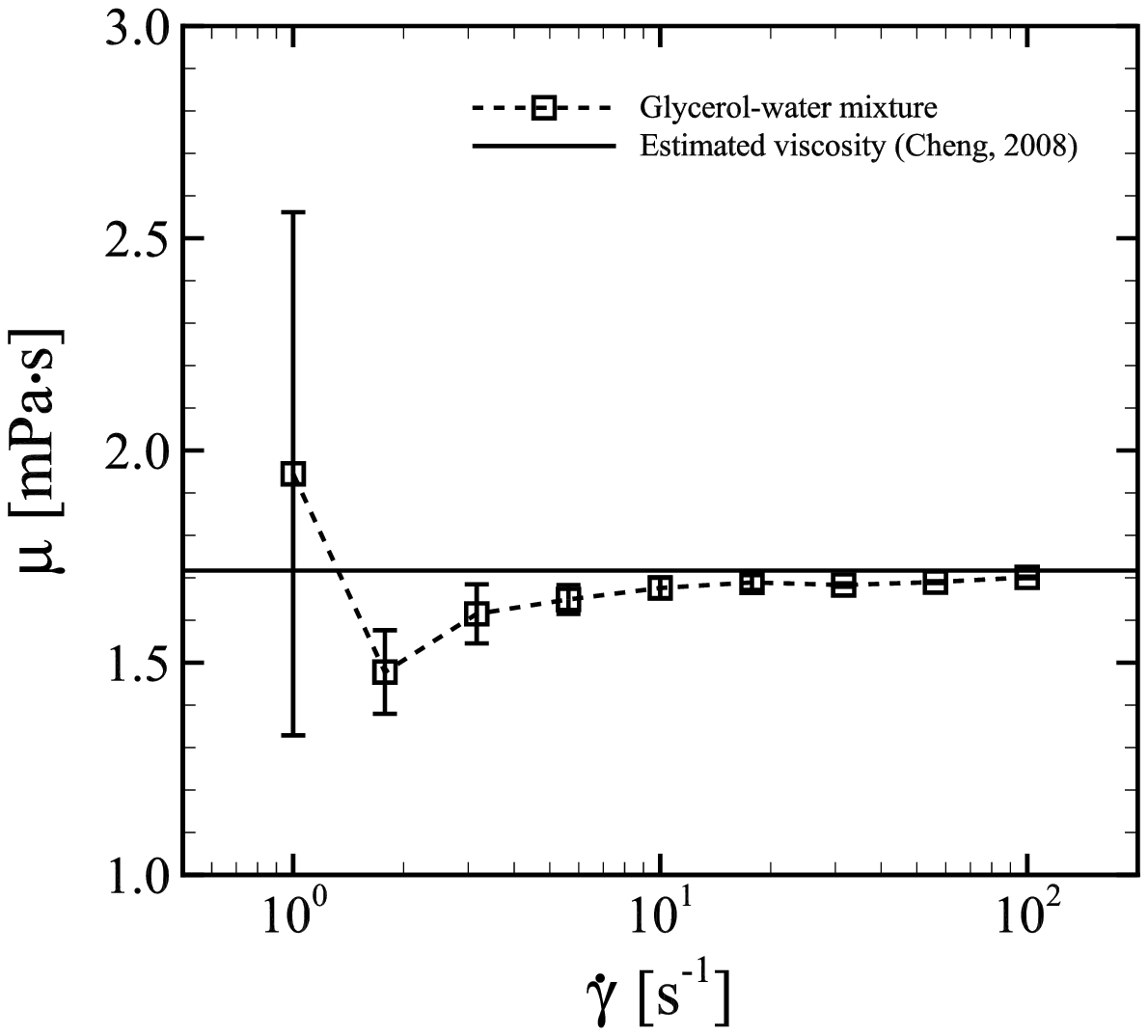}
  \caption{
  Comparison between the shear viscosity of glycerol-water mixture ($22$ $^\circ$C) obtained with experimental measurement and calculation using an empirical expression proposed by~\citet{Cheng2008}.
  The error bars represent SDs (M $\pm$ SD., run cases $n = 4$).
  }
  \label{fig:glycerol_water}
\end{figure}

We measured the shear viscosity of a glycerol-water mixture ($22$ $^\circ$C) as one of Newtonian fluids and compared with that obtained with an empirical expression proposed by~\citet{Cheng2008}.
We confirmed that the relative difference between two values is approximately $1$\% for $\dot\gamma > 10$ s$^{-1}$ as shown in Fig.~\ref{fig:glycerol_water}.
It is known that the rate of change in shear viscosity of glycerol-water mixture is $2$--$3$\% every $1$ $^\circ$C~\cite{Cheng2008}.
Thus, we concluded that the shear viscosity of the plasma shown in Fig.~\ref{fig:shear_viscos}(a) can be evaluated well for $\dot\gamma > 10$ s$^{-1}$.

\section*{References}
\bibliography{reference}

\end{document}